\documentclass[aps,prb,twocolumn,showpacs]{revtex4-1}
\usepackage{amssymb}
\usepackage{graphicx}
\usepackage{amsmath}

\usepackage{times}
\usepackage{color}
\usepackage{setspace}
\usepackage{bm}

\begin{document}
\newcommand {\beq} {\begin{equation}}
\newcommand {\eeq} {\end{equation}}
\newcommand {\bqa} {\begin{eqnarray}}
\newcommand {\eqa} {\end{eqnarray}}
\newcommand {\ba} {\ensuremath{b^\dagger}}
\newcommand {\Ma} {\ensuremath{M^\dagger}}
\newcommand {\psia} {\ensuremath{\psi^\dagger}}
\newcommand {\psita} {\ensuremath{\tilde{\psi}^\dagger}}
\newcommand{\lp} {\ensuremath{{\lambda '}}}
\newcommand{\A} {\ensuremath{{\bf A}}}
\newcommand{\Q} {\ensuremath{{\bf Q}}}
\newcommand{\kk} {\ensuremath{{\bf k}}}
\newcommand{\kp} {\ensuremath{{\bf k'}}}
\newcommand{\rr} {\ensuremath{{\bf r}}}
\newcommand{\rp} {\ensuremath{{\bf r'}}}
\newcommand {\ep} {\ensuremath{\epsilon}}
\newcommand{\nbr} {\ensuremath{\langle ij \rangle}}
\newcommand {\no} {\nonumber}
\newcommand{\up} {\ensuremath{\uparrow}}
\newcommand{\dn} {\ensuremath{\downarrow}}

\begin{abstract}
Interacting bosons with two ``spin'' states in a lattice show novel
superfluid-insulator phase transitions in the presence of spin-orbit
coupling. Depending on the parameter regime, bosons in the
superfluid phase can condense to either a zero momentum state or to
one or multiple states with finite momentum, leading to an
unconventional superfluid phase. We study the response of such a
system to modulation of the optical lattice potential. We show that
the change in momentum distribution after lattice modulation shows
distinct patterns in the Mott and the superfluid phase and these
patterns can be used to detect these phases and the quantum phase
transition between them. Further, the momentum resolved optical
modulation spectroscopy can identify both the gapless (Goldstone)
gapped amplitude (Higgs) mode of the superfluid phase and clearly
distinguish between the superfluid phases with a zero momentum
condensate and a twisted superfluid phase by looking at the location
of these modes in the Brillouin zone. We discuss experiments which
can test our theory.
\end{abstract}
\title{Optical Lattice Modulation Spectroscopy for Spin-orbit Coupled Bosons}
\author{Sangita De Sarkar$^1$, Rajdeep Sensarma$^2$ and K. Sengupta$^1$}
 \affiliation{$1$ Theoretical Physics Department, Indian Association for the Cultivation of
 Science, Jadavpur, Kolkata 700032, India. \\ $2$ Department of Theoretical Physics, Tata Institute of Fundamental
 Research, Mumbai 400005, India.}

\date{\today}

\maketitle
\section{Introduction}
\label{intro}

Ultracold atoms have emerged in recent years as a valuable platform
to study strongly interacting many-particle Hamiltonians relevant to
condensed matter systems, nuclear matter and many other different
fields of physics~\cite{blochreview,essreview}. The unprecedented
control over the Hamiltonian parameters and easy access to strongly
interacting regimes have opened up the possibilities of
systematically studying interacting many body models which have been
used as paradigms to describe a multitude of phenomena from the
realms of condensed matter physics, nuclear physics, astrophysics
and high energy physics. In the arena of condensed matter physics,
many body lattice Hamiltonians like the Bose- and the Fermi-Hubbard
models, which are used as paradigms to study strongly interacting
bosons and fermions, have been experimentally implemented and
studied~\cite{greiner02, jordens08}. These provided a wealth of
information which is relevant to the phenomena of
superfluid-insulator transition~\cite{fisher1} and high temperature
superconductors~\cite{pwascience} respectively. In addition, various
spin-models have also been realized using ultracold
atoms~\cite{trotzky_superex,greinerising}, which will be useful in
study of frustrated magnetic systems and spin
liquids~\cite{optlat_kagome}.

Gauge fields and their interaction with matter are at the heart of
our understanding of the physical phenomena around us. While the
electromagnetic interactions governing most of condensed matter
physics is described by a U(1) Abelian gauge theory, more
complicated non-Abelian versions govern the weak and strong
interactions. In condensed matter systems, the presence of
externally imposed gauge configurations, e.g. a fixed electric or
magnetic field, can lead to interesting and qualitative change in
properties of the system, e.g. in presence of magnetic fields,
vortex lattices can form and melt in a
superconductor~\cite{vortex_sc}, the ground state of the system can
have non-trivial topology and associated quantized conductance in
quantum Hall effect etc~\cite{qhe}. Non-Abelian gauge fields, which
can take the form of spin-orbit coupling~\cite{so_nonabel} is an
essential ingredient in the realization of topological
insulators~\cite{topo_ins} and topological
superconductors~\cite{topo_sc} and plays a crucial role in
understanding novel phenomena like anomalous quantum Hall effect
(AQHE) in systems with strong spin orbit coupling.

Ultracold atoms have been dressed by laser fields~\cite{spielman1},
so that in the lowest manifold of dressed states, the effective
Hamiltonian is that of the bosons/fermions, whose spin and orbital
degrees of freedom are coupled. Alternate
proposals~\cite{sau,galitski_mag_grad} of realizing effective
spin-orbit coupling terms are present in the literature. Recently
time-varying magnetic field gradients have been used to realize
spin-orbit coupling~\cite{mag_grad_expt}. The implementation of
specific types of spin-orbit coupling can lead to interesting phases
of matter like topological insulators and topological
superconductors.  The ability to tune both the effective spin-orbit
coupling and the interaction strength in these systems, which is
very hard to achieve in material based systems, has opened up the
possibility of studying novel Mott insulators and superfluids with
Bose Einstein condensation into states with finite momenta in these
systems~\cite{galitski2008,mandal,ap2014,shizong2014}.

Compared to the wide array of experimental techniques available to
probe material systems, cold atom systems suffer from a paucity of
experimental probes. The main tool of obtaining time of flight
absorption images, which translates to observation of momentum
distribution in lattice systems, is a rather blunt instrument to
differentiate between the myriad phases of matter which can occur in
these systems. Further, the simple time of flight measurement does
not provide any dynamic (energy dependent) information about the
system, which is crucial in understanding its low temperature
properties. A few spectroscopic techniques like rf
spectroscopy~\cite{rf_spec}, Bragg spectroscopy~\cite{Bragg_spec}
and lattice modulation spectroscopy~\cite{sensarma1,sensarma2} are
available to obtain energy resolved information about these systems.
The latter method constitutes an ultracold atom counterpart of
standard angle-resolved photoemission spectroscopy and provides
energy and momentum resolved information regarding the single
particle spectral function of the bosons. This method has been
proposed for single species bosons in the strong-coupling
regime\cite{sensarma1}; however to the best of our knowledge, it has
never been applied to spinor bosonic systems with spin-orbit
coupling. Here, we will focus on the lattice modulation spectroscopy
of such systems.


In this paper, we will consider two component
bosons~\cite{demler1,issacson} in a 2D square optical lattice, which
are interacting with a local Hubbard type interaction. We will
consider ``spin''-orbit
coupling~\cite{grass2011,radic2012,nandini2012,conjun} in these
systems, implemented either by Raman dressing of the atoms or
modulation of magnetic field gradients. We consider the response of
this system to optical lattice modulation spectroscopy and
demonstrate the following. First, we show that optical lattice
modulation spectroscopy can resolve the Mott and the superfluid
phases both by looking at the absence/presence of gapless Goldstone
modes and by identifying the unique pattern of appearance and
disappearance of excitation contours in the Brillouin zone. Second,
we demonstrate that the momentum resolved nature of the optical
modulation spectroscopy can be used to clearly distinguish
superfluids with condensate at zero momentum from those with
condensate at finite momentum. Third, we provide a general theory of
extracting the spectral function of spin-orbit coupled bosons in the
strong coupling regime and near their superfluid-insulator phase
transition by computing their response to the lattice modulation.
Finally, we use the results of this theory to make definite
predictions about the excitation spectrum of these bosons both in
the Mott and the superfluid phases which can tested in realistic
experiments and demonstrate that lattice modulation spectroscopy can
reliably identify and characterize both the gapless (Goldstone) and
the gapped amplitude (Higgs) modes of spin-orbit coupled superfluid
bosons; we note that such an analysis of the properties of
excitation of the SF and Mott phases of these systems is beyond the
scope of standard time-of-flight measurement which can also
distinguish between the Mott and the SF phases.

The plan of the rest of the paper is as follows. In Sec.\
\ref{nonabel}, we discuss the general theory of optical modulation
spectroscopy for spin-orbit coupled bosons. This is followed by
Secs.\ \ref{optmod_mott} and \ref{optmod_sf} where we inspect the
detailed response of the system to optical modulations in the Mott
and the superfluid phases respectively. Finally, we discuss possible
experiments that can be carried out to validate out theory, sum up
our main results, and conclude in Sec.\ \ref{conclusion}.
\begin{figure}[t]
\includegraphics*[width=0.47 \linewidth]{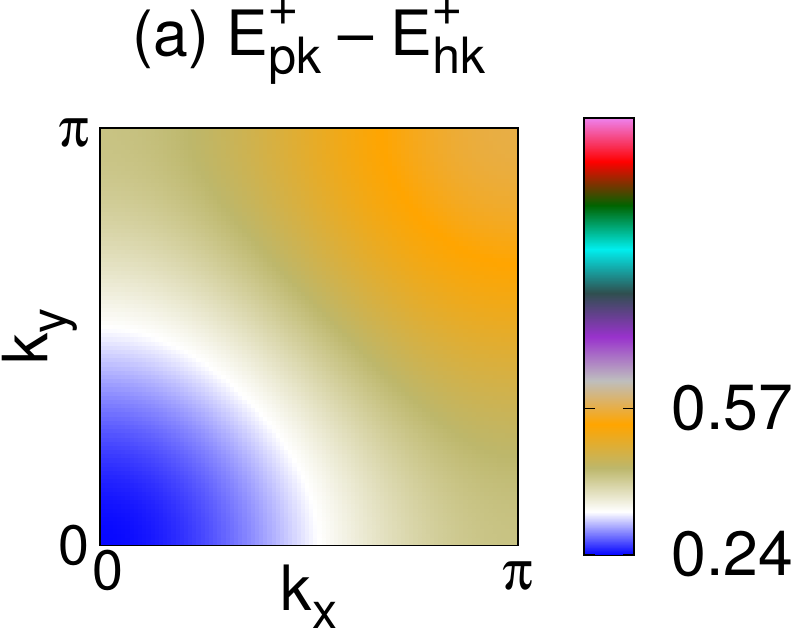}
\includegraphics*[width=0.47 \linewidth]{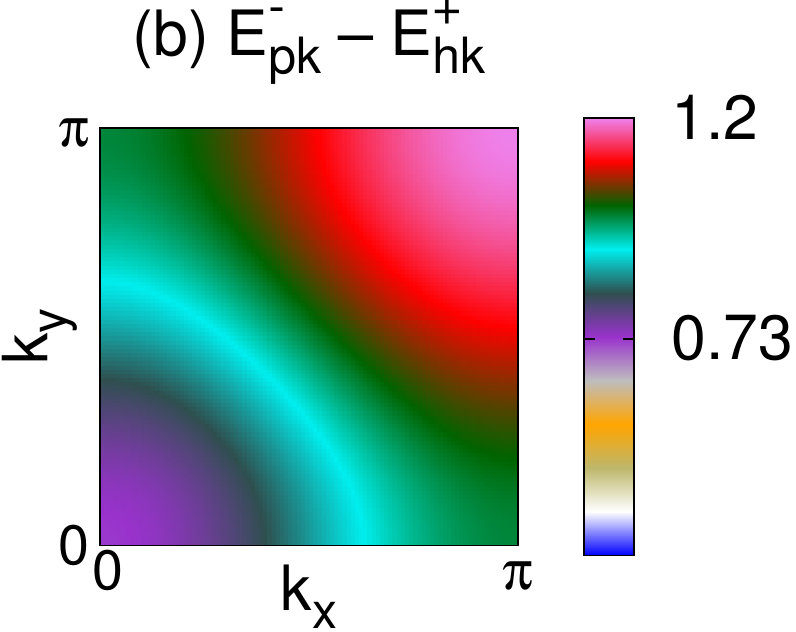}
\includegraphics*[width=0.47 \linewidth]{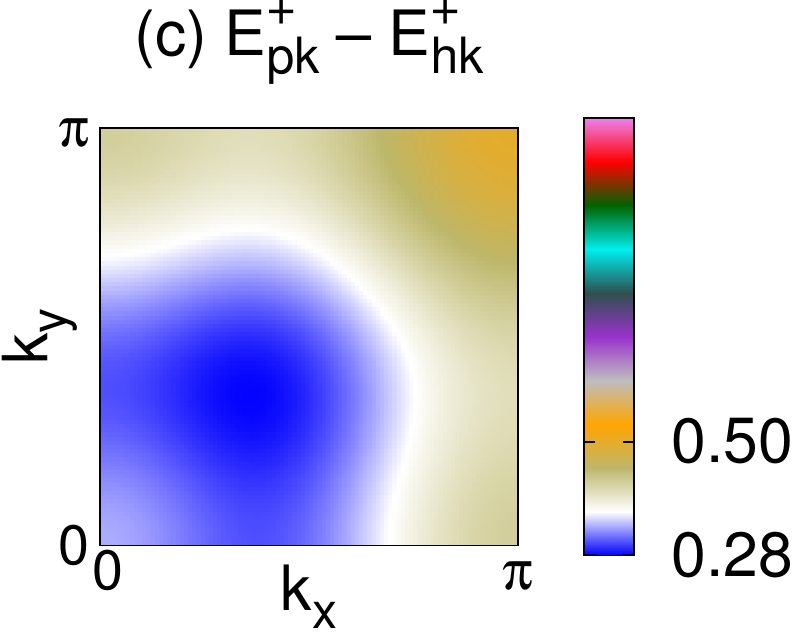}
\includegraphics*[width=0.47 \linewidth]{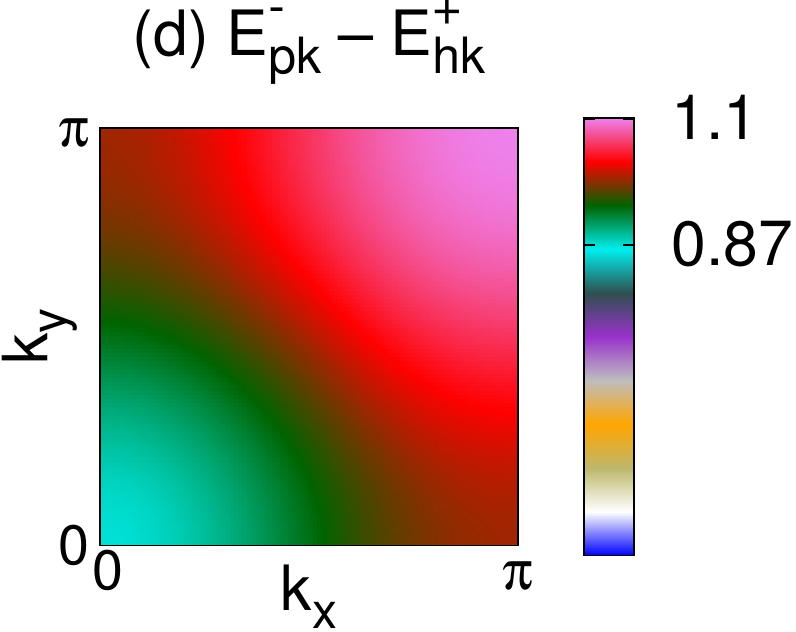}
\caption{Density plots showing (a) intraband and (b) interband
  particle-hole excitation spectrum across the Brillouin zone in the
  Mott phase. The parameters of the system are $J=0.02U$,
  $\gamma=0.01U$, $\mu=0.2U$,$\Omega=0.01U$ and $\zeta=0.4$. The
  intraband excitations have energies between $0.24 U$ to $0.57U$,
  while the interband excitations have energies between $0.73 U$ and $1.2U$. (c)
  Intraband and (d) Interband particle-hole excitations in the Mott
  phase for $J=0.01U$,
  $\gamma=0.04U$, $\mu=0.2U$,$\Omega=0.01U$ and $\zeta=0.4$. In this
  case minimum of intraband excitations are shifted from the zone center
  to $(k_0,k_0)$ with $k_0=1.131 a^{-1}$. The figure shows the contours in the interval
  $0\le k_x,k_y \le \pi$; the contours in the other quadrants can be
  obtained by the reflection of the present figure around $k_x$ and $k_y$ axes and the origin. The
  intraband excitations have energies between $0.28 U$ to $0.5U$,
  while the interband excitations have energies between $0.87 U$ and
  $1.1U$. Note that the color scheme has same
absolute value in all the panels to show which excitations overlap
with each other in energy.} \label{fig1}
\end{figure}

\section {Lattice Modulation Spectroscopy with Non-abelian gauge
  fields }
\label{nonabel}

In a system with multiple species of bosons, Raman lasers can be
used to generate non-abelian gauge fields which couple the different
species of bosons. In most experimental situations, the different
boson species are actually different hyperfine state of the same
bosonic species allowing one to treat them as a system of
multi-component bosons. In fact, if one can trap two bosonic states
it can be considered as an effective pseudo-spin $1/2$ system, which
is however made of bosonic atoms. A particularly interesting
non-abelian gauge field configuration is the Rashba spin-orbit
coupling in the two component bosonic system. There are various
proposals to implement the Rashba spin-orbit coupling, although
currently experiments have focussed on the more easily implementable
configuration of equal Rashba and Dresselhouse coupling, which leads
to spins coupling to momenta in a particular direction. Interacting
bosons with Rashba spin-orbit coupling shows interesting chiral Mott
and superfluid phases as various parameters are tuned in the
Hamiltonian \cite{xu1,conjun}.

The Hamiltonian for two-species bosons in a square optical lattice in
the presence of Rashba spin-orbit coupling term can be written as \cite{issacson,demler1,haldane1}
\bqa
H_0 &=& \sum_{i \sigma} [(-\Omega\sigma^z- \mu )n_{i\sigma} + \frac{U}{2}  n_{i\sigma}( n_{i\sigma}-1)] +\zeta U  n_{i\sigma}  n_{i\bar{\sigma}} \no\\
&& -  J\sum_{\langle ij\rangle \sigma} b_{i\sigma}^{\dagger} b_{j\sigma}+i \gamma \sum_{\langle ij\rangle} {\hat \Psi}_i^{\dagger}
{\hat z} \cdot \left(
 {\vec \sigma} \times {\vec d}_{ij} \right) {\hat \Psi}_j
\label{ham1} \eqa Here $b_{i\sigma}$ annihilates a boson of {\it
spin} $\sigma= \up,\dn$ on the ${\rm i^{th}} $ site, $n_{i\sigma}
=b_{i\sigma}^{\dagger} b_{i\sigma}$ is the number of $\sigma$
bosons, $U(\zeta U)$ is the intra-(inter-)species interaction
strength between the bosons, and $J$ denotes the nearest neighbor
hopping amplitude. Here $\mu$ and $\Omega$ are the species
independent and the species dependent chemical potentials;  the
latter acts as an effective Zeeman magnetic field for these bosons.
The last term represents the lattice analogue of the Rashba
spin-orbit coupling generated by the Raman lasers
\cite{haldane1,ftnote2}, with a coupling constant $\gamma$. Here,
${\vec d}_{ij}$ is unit vector along the $x-y$ plane between the
neighboring sites $i$ and $j$, $\vec{\sigma}$ is the vector of Pauli
matrices, and $\hat {\Psi}_i= (b_{i\up},b_{i\dn})$ is the
two-component boson field.

The non-interacting part of the Hamiltonian is given by
\begin{eqnarray}
H_K &=& \sum_k \Psi^\dagger_k
[\hat{\Lambda}(k)-\Omega\sigma^z-\mu]\Psi_k
\nonumber\\
\hat{\Lambda}(k) &=& \epsilon_k 1+\gamma_k\sigma^++\gamma^\ast_k
\sigma^- \label{lamdef}
\end{eqnarray}
 where $k \equiv {\bf k}=(k_x,k_y)$ is the 2D quasi-momentum,
$\epsilon_k =-2J(\cos k_x +\cos k_y)$ and $\gamma_k =-2\gamma i
(\sin k_x -i\sin k_y)$. This can be diagonalized to obtain chiral
bands touching each other at the zone center for $\Omega=0$, while a
finite $\Omega$ opens up a gap between the two bands. The ratio
$\gamma/J$ controls the bare dispersion including the location of
the band-minima. As $\gamma/J$ becomes larger the location of the
minimum of the lowest band shifts away from the zone center,
$[0,0]$.
\begin{figure}[t]
\includegraphics*[width=0.47 \linewidth]{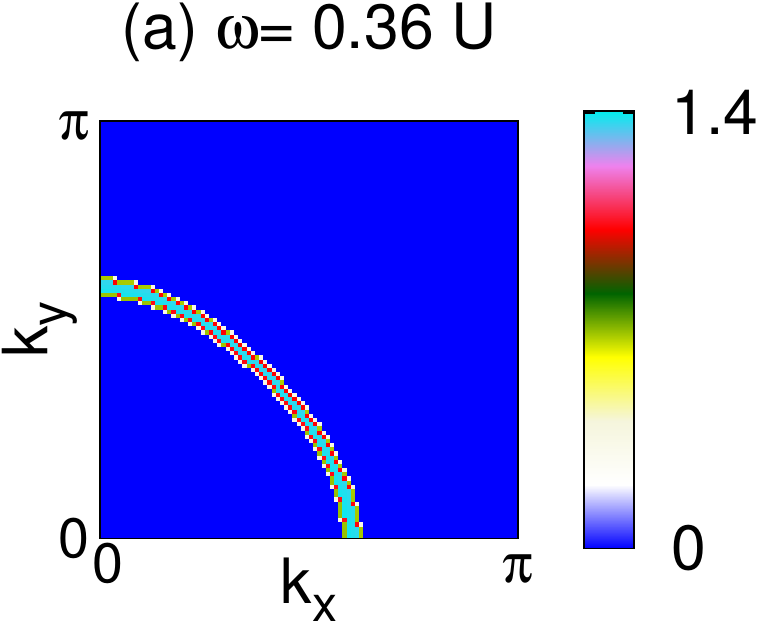}
\includegraphics*[width=0.47 \linewidth]{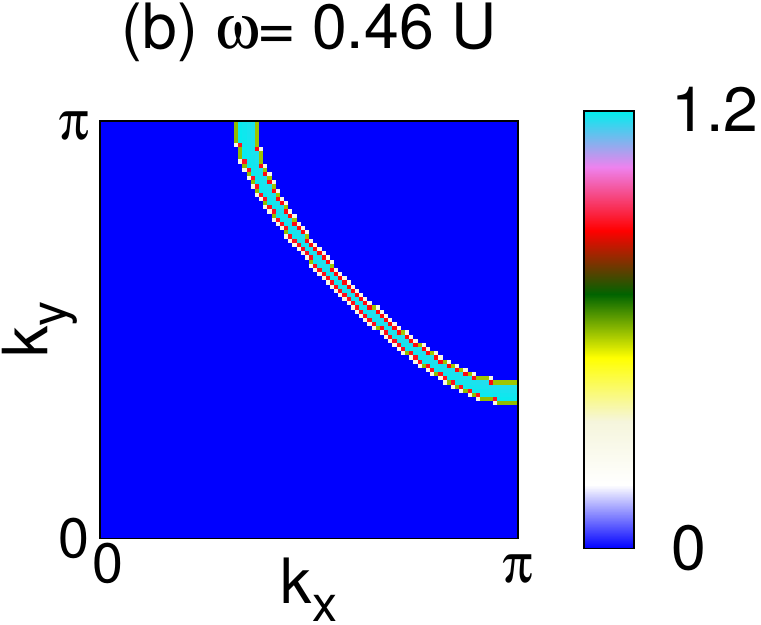}
\includegraphics*[width=0.47 \linewidth]{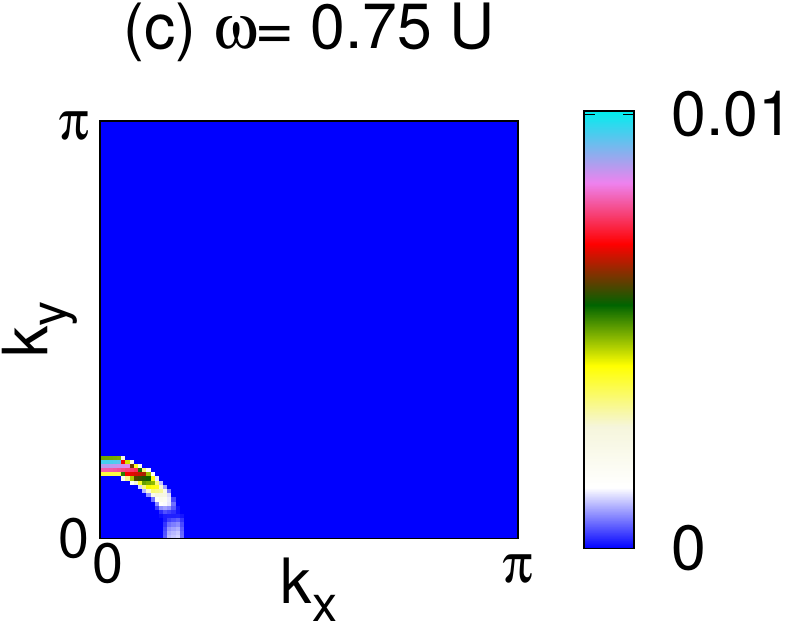}
\includegraphics*[width=0.47 \linewidth]{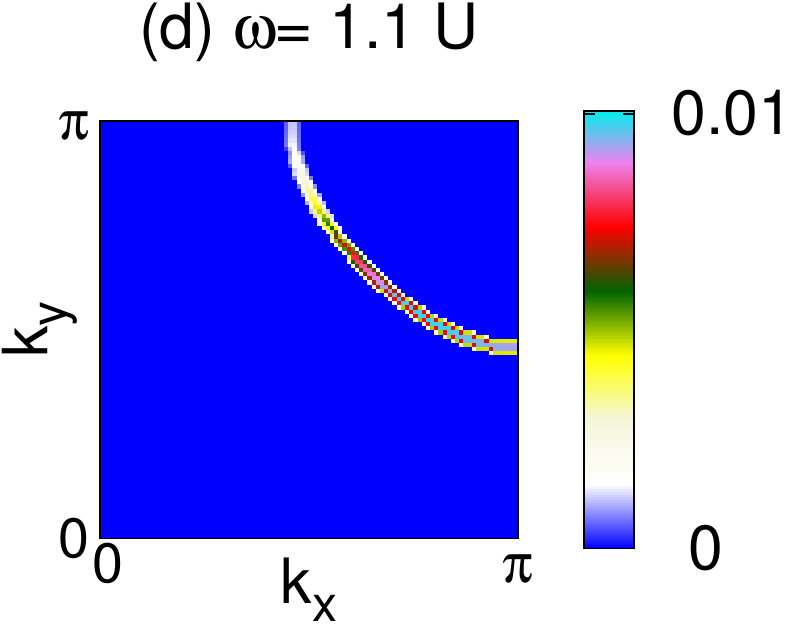}
\caption{Lattice Modulation response in the Mott phase for different
   modulation frequencies. The parameters of the system are $J=0.02U$,
   $\gamma=0.01U$, $\mu=0.2U$,$\Omega=0.01U$ and $\zeta=0.4$. As the
   frequency of modulation crosses the Mott gap, a contour of
   intraband excitations are seen to disperse across the Brillouin
   zone ( a and b). Then, as the frequency crosses the interband
   threshold, another contour of interband excitations disperse across
 the Brillouin zone (c and d). Note that the spectral weight of the interband transitions
 are much smaller than the weights of other transitions.}
 \label{fig2}
 \end{figure}
\begin{figure}[t]
\includegraphics*[width=0.47 \linewidth]{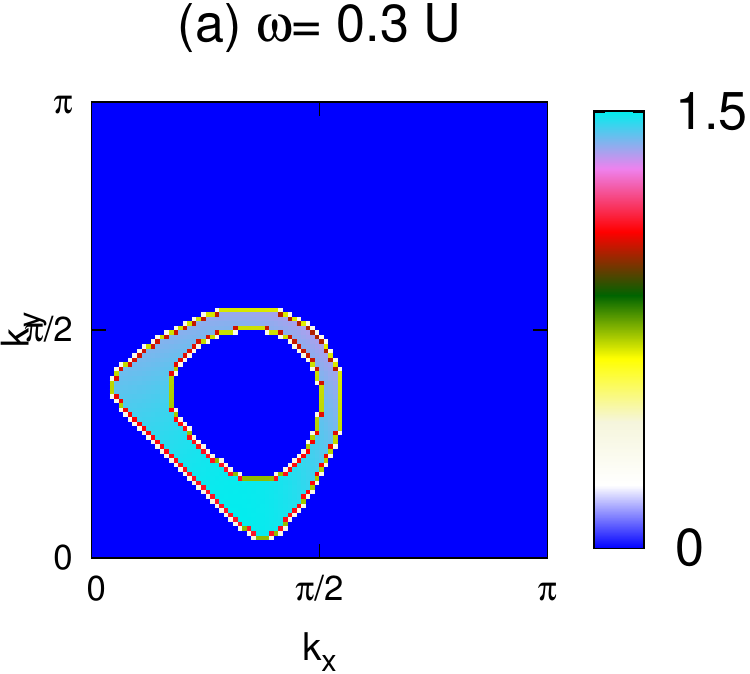}
\includegraphics*[width=0.47 \linewidth]{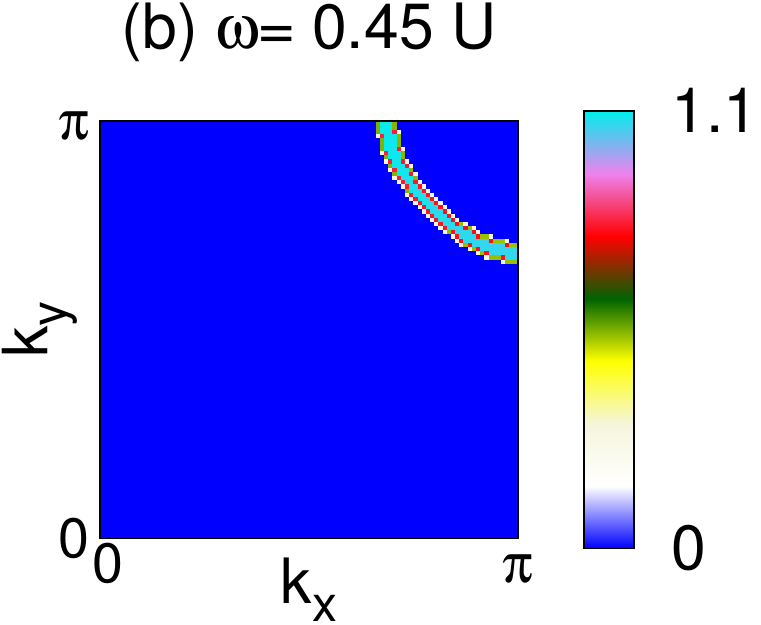}
\includegraphics*[width=0.47 \linewidth]{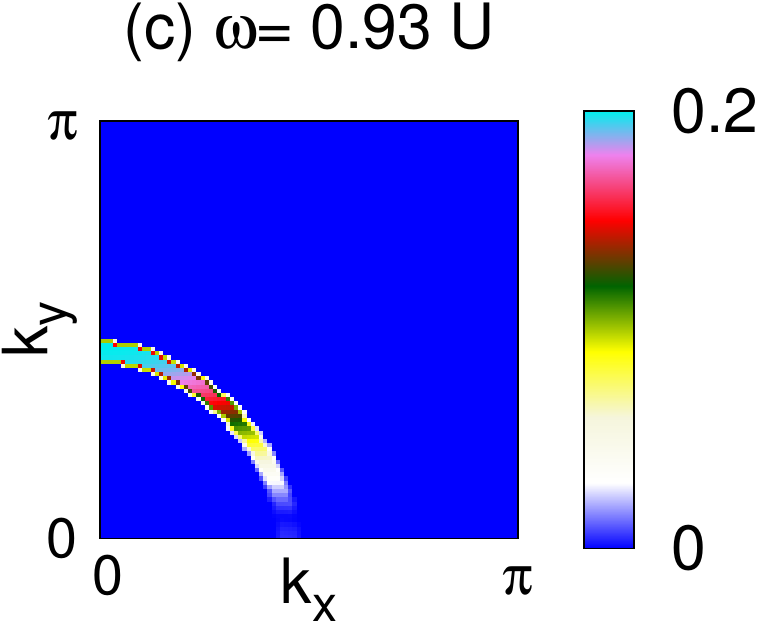}
\includegraphics*[width=0.47 \linewidth]{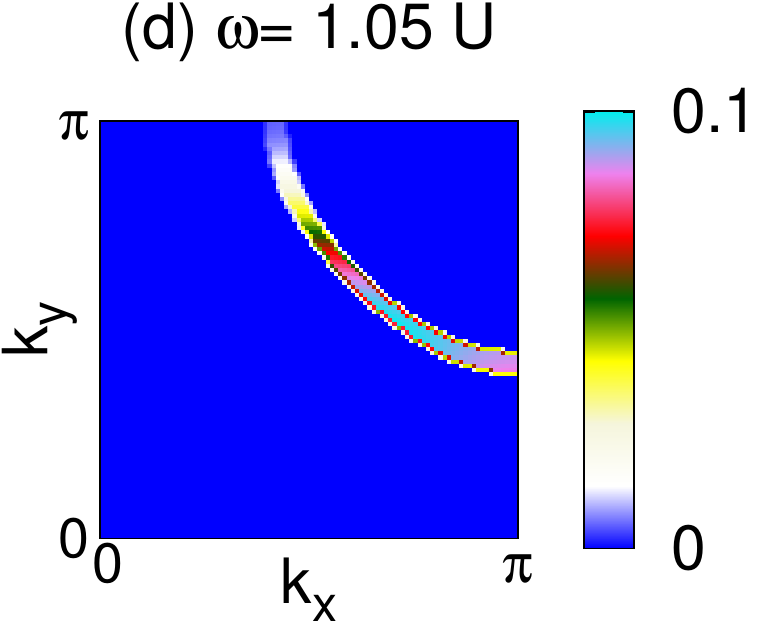}
\caption{Lattice Modulation response in the Mott phase for different
  modulation frequencies. The parameters of the system are $J=0.01U$,
  $\gamma=0.04U$, $\mu=0.2U$,$\Omega=0.01U$ and $\zeta=0.4$. In this
  case minimum of intraband excitations are shifted from the zone center to $(\pm k_0,\pm k_0)$
  with $k_0=1.131 a^{-1}$. The figure shows the contours in the interval
  $0\le k_x,k_y \le \pi$; the contours in the other quadrants can be
  obtained by the reflection of the present figure around $k_x$ and $k_y$ axes and the origin. As the
   frequency of modulation crosses the Mott gap, a contour of
   intraband excitations are seen to disperse across the Brillouin
   zone starting from a circle around $[k_0,k_0]$ ( a and b). Then, as the frequency crosses the interband
   threshold, another contour of interband excitations disperse across
 the Brillouin zone starting from the zone center (c and d).}
\label{fig3}
\end{figure}

The spin-orbit coupled bosons undergo a Mott insulator to superfluid
quantum phase transition as a function of $J/U$ and
$\gamma/U$\cite{mandal,grass2011}. In the strongly interacting
limit, each site has exactly $n^0_\sigma$ bosons of spin $\sigma$,
with the value of $n^0_\sigma$ decided by $\mu$ and $\Omega$. The
system is thus a Mott insulator with no number or spin fluctuations.
As $J/U$ or $\gamma/U$ increases, the system undergoes a phase
transition to a state with delocalized bosons. This is the
superfluid state with phase coherent Bose condensate and gapless
Goldstone excitations due to broken $U(1)$ symmetry. If the
transition takes place when $\gamma/J$ is large, the minimum of  the
effective dispersion occurs at $[\pm k_0,\pm k_0]$, and the bosons
condense into these finite momentum states, leading to a twisted
superfluid phase \cite{mandal}. At low values of $\gamma/J$ one
recovers the standard superfluid with a condensate at zero momentum.
Thus this system shows two remarkable qualitative changes : (a) a
superfluid-Mott insulator transition as a function of $J/U$ and
$\gamma/U$ and (b) a change from a standard superfluid to a twisted
superfluid as a function of $\gamma/J$. As we will show, lattice
modulation spectroscopy can distinguish both these phenomena and
hence provide us with a wealth of information about this system.

The optical lattice modulation spectroscopy protocol that we propose
consists of the following steps: (a) The optical lattice potential
is weakly modulated with an a.c. field on top of the static field
that forms the lattice in the original system, with the Raman fields
and the trapping potential turned on. This leads to a modulation of
the hopping parameter and the spin-orbit coupling. (b)  The
modulation is turned off after some time, making sure that the
system is still in perturbative regime. At the same time, the Raman
lasers, the optical lattice lasers and the trapping potential is
also turned off and the system undergoes ballistic expansion, from
which the (spin-resolved) momentum distribution of the system right
after the modulation can be measured. (c) The change in the momentum
distribution (from the unperturbed/ unmodulated system) will provide
us with information about the spectrum and spectral weight of one
particle excitations in these systems~\cite{sensarma1}.
\begin{figure*}[t]
\includegraphics[width=0.33 \linewidth]{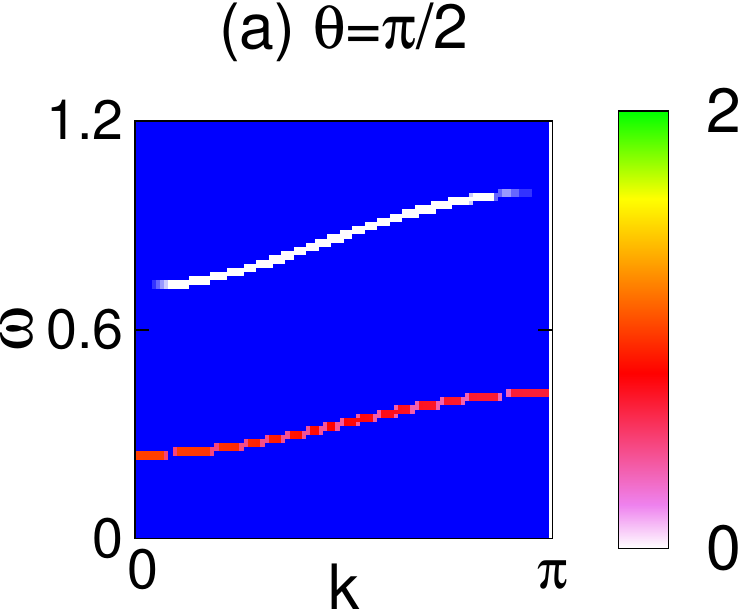}
\includegraphics[width=0.33 \linewidth]{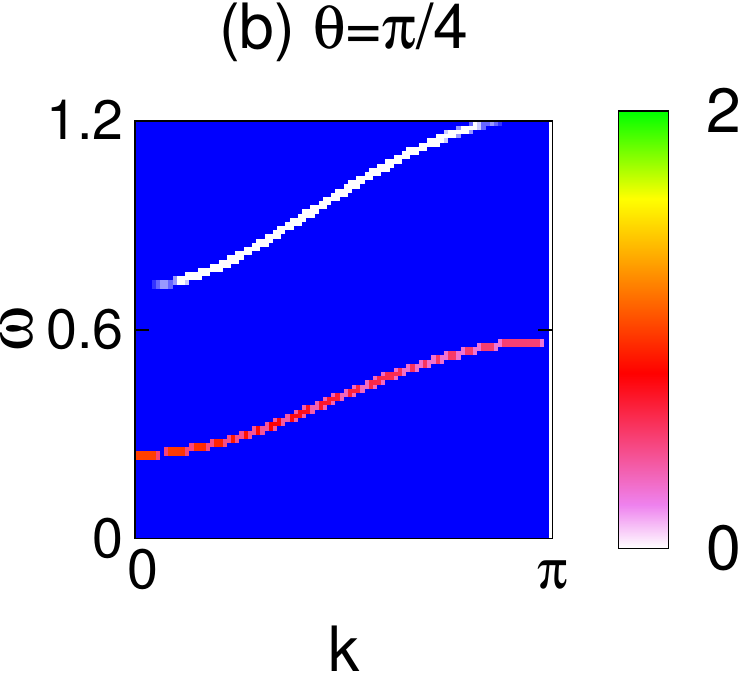}
\includegraphics[width=0.33 \linewidth]{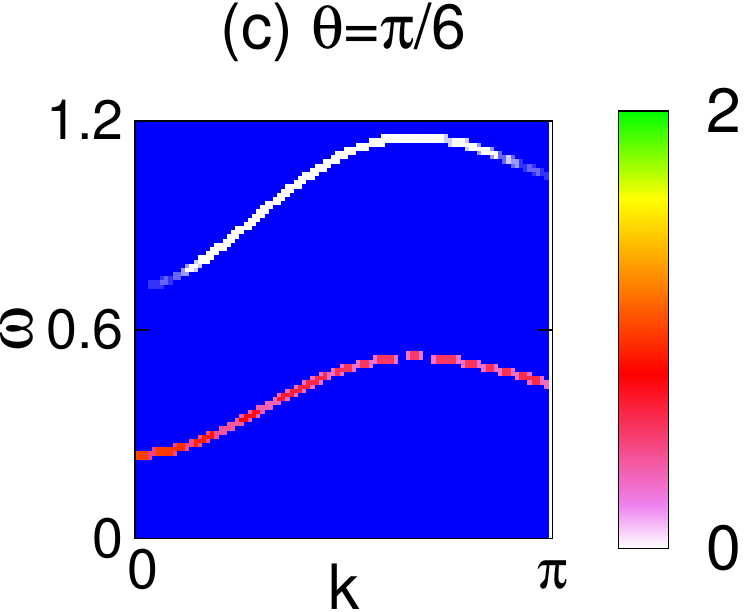}
\includegraphics[width=0.33 \linewidth]{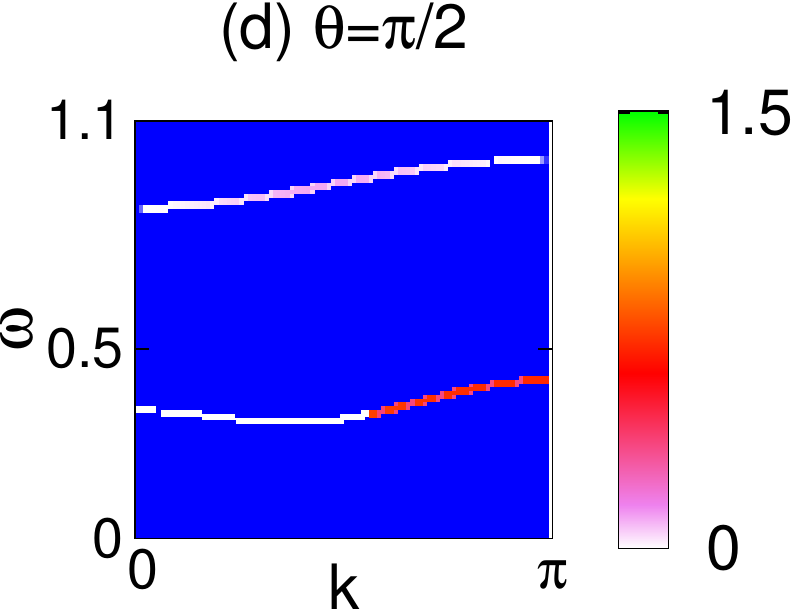}
\includegraphics[width=0.33 \linewidth]{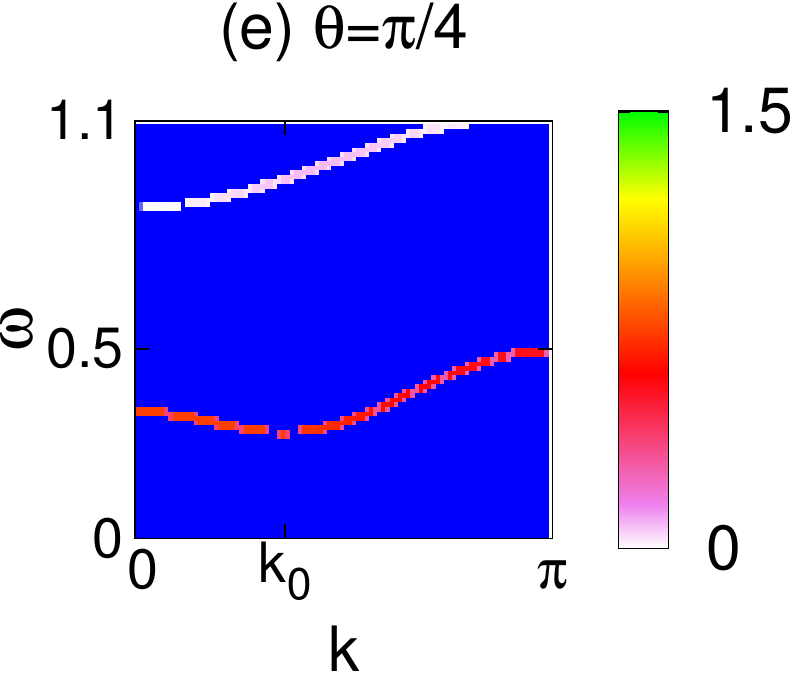}
\includegraphics[width=0.33 \linewidth]{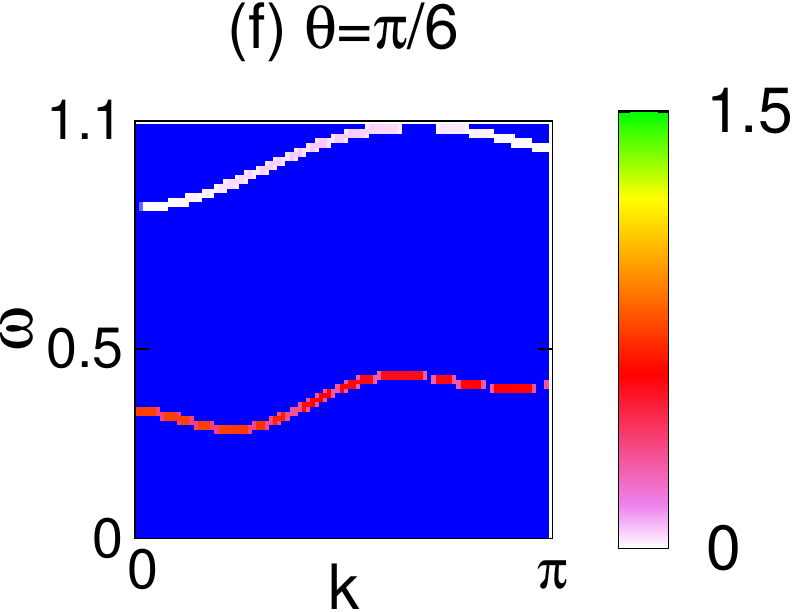}
\caption{Lattice modulation response in the Mott phase with the
variation in frequency for various cuts in the Brillouin zone going
radially outwards at an angle $\theta$ with the $k_x$ axis. In (a) -
(c) the parameters of the system are $J=0.02U$,
  $\gamma=0.01U$, $\mu=0.2U$,$\Omega=0.01U$ and $\zeta=0.4$ with the
  band minimum at the zone center. (a) $\theta=\pi/2$, (b)
  $\theta=\pi/4$ and (c) $\theta=\pi/6$. In (d) - (f) the parameters
  of the system are  $J=0.01U$,
  $\gamma=0.04U$, $\mu=0.2U$,$\Omega=0.01U$ and $\zeta=0.4$. In this
  case minimum of intraband excitations are shifted from the zone center
  to $(k_0,k_0)$ with $k_0=1.131 a^{-1}$.
  (d) $\theta=\pi/2$, (e) $\theta=\pi/4$ and (f) $\theta=\pi/6$.} \label{fig8}
\end{figure*}

In experiments, the lattice Hamiltonian Eq.~\ref{ham1} is
implemented by putting a system of bosons, characterized by mass
$m$, a continuum spin-orbit coupling $\gamma_c$ and an effective
magnetic field $h$  generated by the Raman lasers, under a periodic
optical potential $V_0[\cos^2 (x/a) +\cos^2 (y/a)]$, where
$a=\lambda_{op}/(2\pi)$ is a transverse confinement scale used to
construct the 2D square lattice, and $\lambda_{op}$ is the
wavelength of the laser forming the optical lattice. The interaction
between these bosons in continuum are characterized by a
intra-species scattering length $a_s$ and an inter-species
scattering length $\zeta a_s$. In the limit of a deep lattice, the
continuum parameters are related to the lattice parameters by
\begin{eqnarray}
U &=& 4\sqrt{V_0E_R}a_s/a, \quad J=\sqrt{V_0E_R}
e^{-\frac{1}{4}\sqrt{V_0/E_R}} \nonumber\\
\gamma &=& \gamma_c\frac{\sqrt{2mV_0}}{\hbar}
e^{-\frac{1}{4}\sqrt{V_0/E_R}}, \label{paramtr1}
\end{eqnarray}
where $E_R=\hbar^2/2m\lambda_{op}^2$ is the recoil energy. It is
evident that $J$ and $\gamma$ both depend exponentially on the
lattice depth $V_0$, while $U$ has only polynomial dependence on
lattice height. Thus, an ac modulation put on the optical lattice
depth $V_0(t)=V_0+\delta V \cos\omega t$ will lead to simultaneous
modulation of the hopping parameter, the spin orbit coupling as well
as the interaction parameter $U$. Using $U$ as an overall scale for
the problem, the perturbation Hamiltonian, to linear order in
variations of the optical lattice potential, can be written as
\begin{eqnarray}
\nonumber H_1&=& \delta U H_0 - U\delta\left(\frac{J}{U}\right)\sum_{\langle
  ij\rangle \sigma} b_{i\sigma}^{\dagger} b_{j\sigma}\\
& &+i U\delta\left(\frac{\gamma}{U}\right) \sum_{\langle ij\rangle} {\hat \Psi}_i^{\dagger}
{\hat z} \cdot \left(
 {\vec \sigma} \times {\vec d}_{ij} \right) {\hat \Psi}_j
\end{eqnarray}
where the first term, proportional to the variation of $U,$ commutes
with the unperturbed Hamiltonian and hence does not create
excitations. This term can thus be neglected as far as modulation
spectroscopy is concerned. In this case, the perturbation consists
of modulating the hopping and spin-orbit terms with amplitudes
\cite{ftnote1}
\begin{eqnarray}
\lambda &=& \delta J/J=\delta \gamma/\gamma=\frac{1}{8}(\delta
V/\sqrt{V_0E_R}) \label{modparamtr1}
\end{eqnarray}
The perturbation Hamiltonian can then be written
as
\beq
 H_1(t)= \lambda \cos \omega t \sum_k \Psi^\dagger_k
\hat{\Lambda}(k)\Psi_k.
\eeq
In linear response regime, the momentum distribution right after the
modulation is turned off oscillates with the frequency of the
perturbation \cite{sensarma1}
\beq \delta n_{\sigma,k}(t)=\delta n^{(1)}(\sigma,k,\omega)\cos
\omega t +\delta n ^{(2)}(\sigma,k,\omega)\sin \omega t \eeq
The out-of phase response contains the information about excitation
spectrum of the unperturbed system. In the next section, we will
define the response function $\Pi_\sigma(k,i\omega)$, where
$i\omega$ is the Matsubara frequency,  and relate this to the single
particle Green's function of the bosons and hence to the excitation
spectra. The imaginary part of $\Pi_\sigma(k,\omega+i0^+)$ then
measures the amplitude of the out of phase modulation of the spin
dependent momentum distribution due to the perturbation. The
structure of the response is qualitatively different in the Mott and
superfluid phase owing to the broken $U(1)$ symmetry in the
superfluid phase and associated anomalous propagators. Hence we will
treat the response in the Mott and superfluid phase separately.

\section {Response in the Mott phase }
\label{optmod_mott}

In the strongly interacting limit, the system is in an
incompressible Mott insulating state with gapped excitation
spectrum. In the atomic limit, ($\gamma=J=0$), the ground state has
a fixed number of bosons of each spin, $n^0_\sigma$ at all the
sites. In general, $n^0_\sigma$ is determined by the values of $\mu$
and $\Omega$. We will restrict our analysis to the case where
$n^0_\up=1$ and $n^0_\dn=0$, but the analysis can be easily extended
to arbitrary integer values of $n^0_\sigma$.  The single particle
Green's function in the Mott phase is a $2\times 2$ matrix in the
spin space, in terms of which the response function is given by
 %
 \bqa
\label{piknonabel} \no
\Pi_\sigma(k,i\omega_n)&=&\frac{\lambda}{\beta}\sum_{\omega_l}
[\hat{G}(k,i\omega_l)\hat{\Lambda}(k)\hat{G}(k,i\omega_l+i\omega_n)]_{\sigma\sigma}\\
& &+(\omega_n\rightarrow -\omega_n)
\eqa
%
The boson Green's function can be worked out in a strong coupling
expansion around the localized atomic limit
\cite{sengupta1,mandal,grass2011}. In the Mott phase, it is given by
\begin{eqnarray}
G^{-1}(k,i\omega_n) &=& \left(\begin{array}{cc}
F_1(i\omega_n)-\epsilon_k  &-\gamma_k\\
-\gamma^\ast_k & F_2(i\omega_n)-\epsilon_k
\end{array}\right) \nonumber\\
F_1(i\omega_n) &=& i\omega_n+E_0-
2U+2U^2/(i\omega_n+E_0+U),\nonumber\\
F_2(i\omega_n) &=& i\omega_n-E_1 \label{eq:soginv}
\end{eqnarray}
where $E_0=\mu+\Omega$ and $E_1=\Omega+\zeta U-\mu$.

It is evident that this Green's function is not diagonal in the
basis of non-interacting bands, as the atomic limit local propagator
is different for the 2 spin species. This can be traced to the fact
that in the atomic limit ($J=0$, $\gamma=0$) the $\Omega$ term lifts
the degeneracy between the spin states and as a result one obtains a
polarized Mott state which is $n_0=1$ for the $\up$ spins and
$n_0=0$ (vacuum) for the $\dn$ spins. The Green's function can,
however, be diagonalized to obtain
\beq
G_D(k,i\omega_n)=\left(\begin{array}{cc}
\frac{1}{\zeta_-(k,i\omega_n)}  &0\\
0 & \frac{1}{\zeta_+(k,i\omega_n)}
\end{array}\right)
\label{eq:sog}
\eeq
\begin{figure}
\includegraphics*[width=0.47 \linewidth]{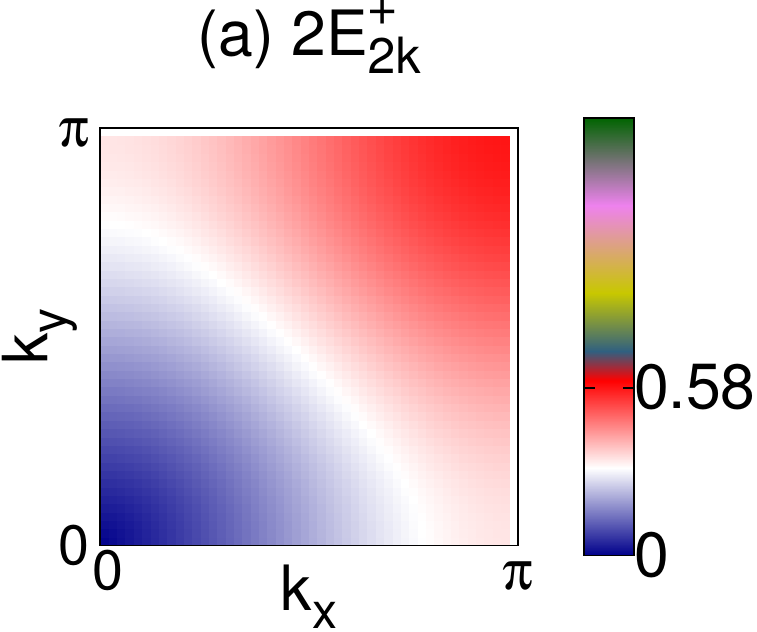}
\includegraphics*[width=0.47 \linewidth]{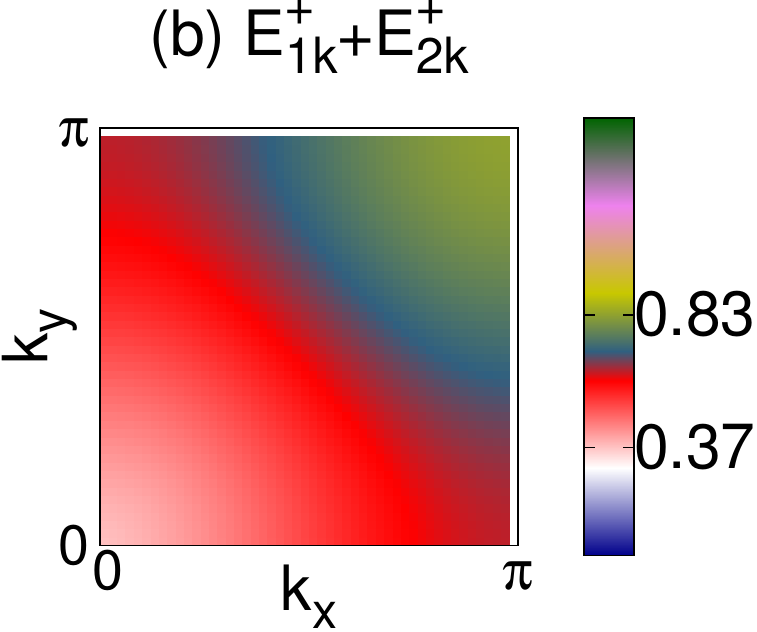}
\includegraphics*[width=0.47\linewidth]{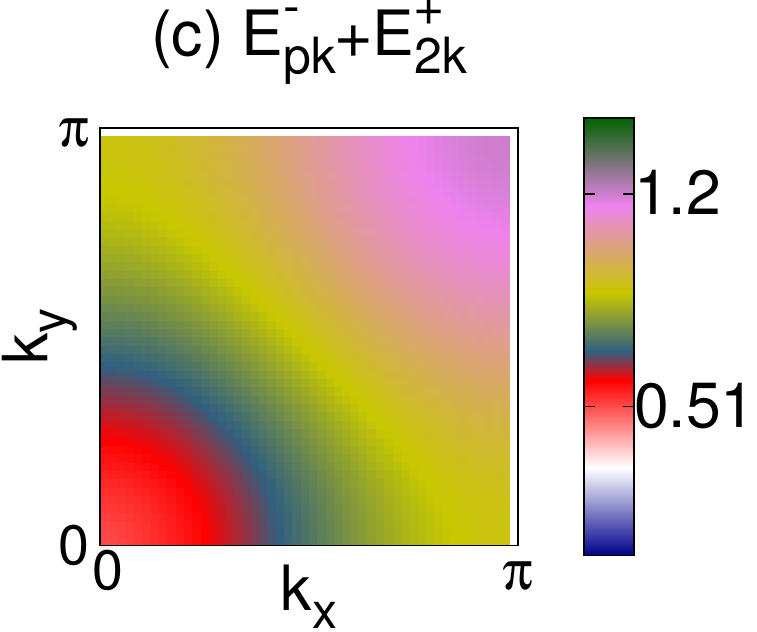}
\includegraphics*[width=0.47 \linewidth]{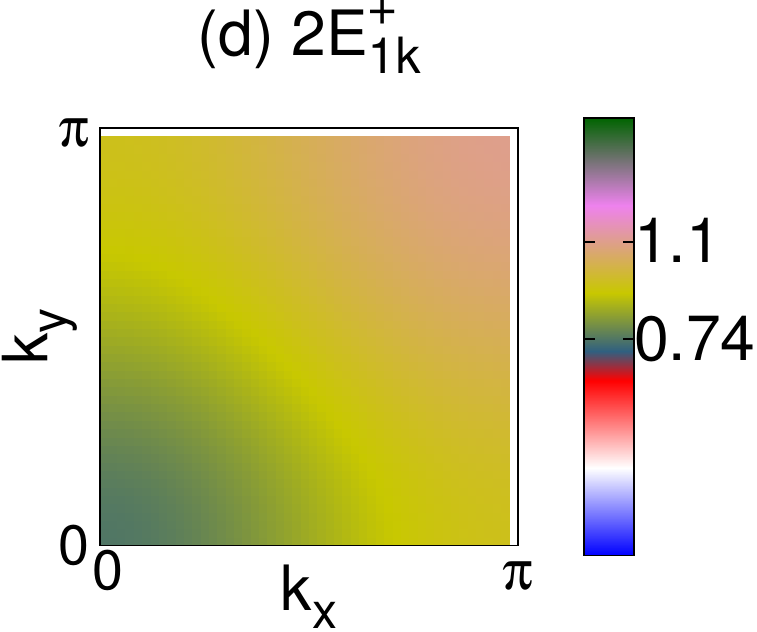}
\includegraphics*[width=0.47\linewidth]{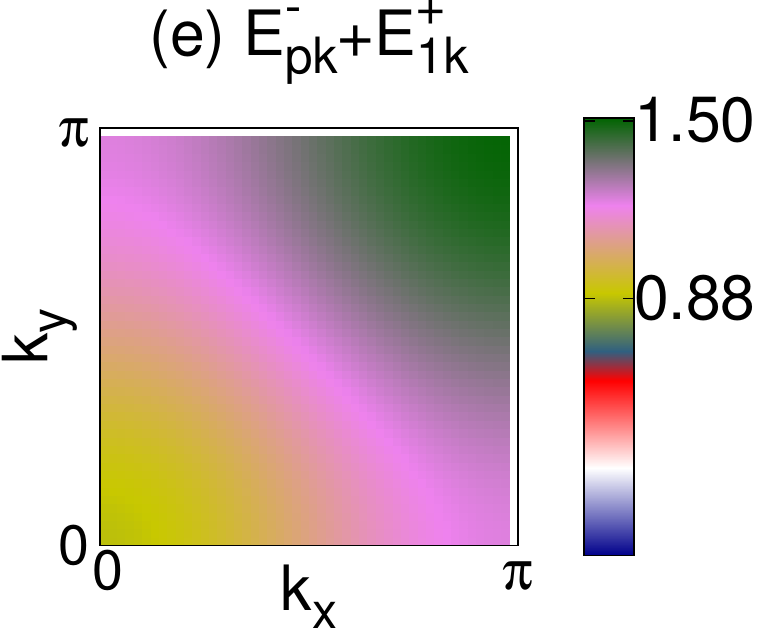}
\caption{Excitation spectrum in the superfluid phase. The parameters
of the system are $J=0.03U$,
  $\gamma=0.01U$, $\mu=0.35U$, $\Omega=0.01U$ and $\zeta=0.4$. In this
  case the BEC is formed at the zone center. The excitations are (a) two
phase modes (b) one phase and one amplitude mode (c) one phase mode
and one interband transition (d) two amplitude modes and (e) an
amplitude mode and an interband transition. Note that the color
scheme has same absolute value in all panels to show which
excitations overlap with each other in energy.} \label{fig4}
\end{figure}
where the diagonal components are given by
\beq
\zeta_{\pm}(k,i\omega_n)=F_+(i\omega_n)-\epsilon_k \pm
(F^2_-(i\omega_n)+|\gamma_k|^2)^{1/2},
\label{eq:zeta}
\eeq
 and
$F_{\pm}(i\omega_n)=(1/2)[F_1(i\omega_n)\pm F_2(i\omega_n)]$. The basis transform,
which diagonalizes the Green's function is given by
\begin{eqnarray}
M(k,i\omega_n) &=& \left(\begin{array}{cc}
u(k,i\omega_n)  &v(k,i\omega_n)\\
-v^\ast(k,i\omega_n) & u^\ast(k,i\omega_n)
\end{array}\right) \nonumber\\
|u(k,i\omega_n)|^2&=& \gamma_k^2/N(k,i\omega_n)
\nonumber\\
v(k,i\omega_n) &=& \gamma_k^{-1} u^{\ast}(k,i\omega_n)[
F_1(i\omega_n)-\epsilon_k-\zeta_{+}(k,i\omega_n)],
\nonumber\\
N(k,i\omega_n)&=&\sqrt{[F_1(i\omega_n)-\epsilon_k-\zeta_{+}(k,i\omega_n)]^2+\gamma_k^2}.
\label{eq:somat} \end{eqnarray}
Note that $|u(k,i\omega_n)|^2+|v(k,i\omega_n)|^2=1$.

It is evident from the above expressions that even in the transformed
basis, the diagonal Green's function (Eq. ~\ref{eq:sog}) will have a complicated frequency
dependence. However, we are only interested in the out of phase
response of the system to the external perturbation, which depends
solely on the imaginary part of the Green's function, analytically
continued to the real frequency domain ($i\omega_n \rightarrow \omega +i0^+$). It can be easily
shown that the retarded Green's function only has simple poles, and so, for the
purpose of calculating the out of phase response, the complicated
expression of Eq.~\ref{eq:sog} can be replaced by the simpler form
\beq
G_D(k,i\omega_n)\simeq\left(\begin{array}{cc}
  \frac{z^-_{pk}}{i\omega_n-E^-_{pk}}  & 0\\
0 &  \frac{z^+_{pk}}{i\omega_n-E^+_{pk}}+\frac{z^+_{hk}}{i\omega_n-E^+_{hk}}
\end{array}\right)
\label{eq:sogapp}
\eeq
where $\omega=E^+_{p(h)k}$ are the two zeroes of $\zeta^+(k,\omega)$
and correspond to the particle and hole excitations of the $+$ band,
while $\omega=E^-_{pk}$ is the zero of $\zeta^-(k,\omega)$.  The
residues $z$ are given by the $\omega$ derivative of the components
of the Green's function, Eq.~\ref{eq:sog}, evaluated at the
corresponding poles. The three poles have a relatively simple
explanation in terms of the particle and hole excitations of a Mott
insulator. In absence of the spin-orbit coupling ($\gamma=0$), the
$\up$ spins form a $n_0=1$ Mott insulator and has the corresponding
particle and hole excitations. This corresponds to the limiting form
of $E^+_{p(h)k}$ as $\gamma \rightarrow 0$. The $\dn$ spins ,
however, form a $n_0=0$ Mott insulator, and hence has only particle
excitations. This is the limiting form of $E^-_{pk}$ in the limit
$\gamma \rightarrow 0$. In presence of spin-orbit coupling, the spin
states get mixed, but this three pole structure of the Green's
function persists. We note that the Green's function in
Eq.~\ref{eq:sogapp} has the same imaginary part as the actual
Green's function (Eq. ~\ref{eq:sog}) and hence there is no
approximation made in the above replacement, as far as
computation of the out of phase response of the system is
concerned.

The response of the system can now be calculated in terms of the
diagonal Green's function as
\begin{widetext}
\bqa \no \Pi_\sigma(k, i\omega_n)&=&\frac{\lambda}{\beta}\sum_{
i\omega_l}\left[ M(k,i\omega_l)G_D(k,i\omega_l)M^{-1}(k,i\omega_l)
\hat{\Lambda}(k)M(k,i\omega_l+i\omega_n)G_D(k,i\omega_l+i\omega_n)M^{-1}(k,i\omega_l+i\omega_n)\right]_{\sigma\sigma}\\
& &+\omega_n \rightarrow -\omega_n, \eqa
\end{widetext}
 where ${\hat \Lambda}(k)$ is given by Eq.\ \ref{lamdef}. Note
that the transformation matrices are themselves function of the
Matsubara frequencies $i\omega_l$ and $i\omega_n$. However, it can
be easily seen that the transformation matrix elements, analytically
continued to the real frequency domain ($i\omega_l \rightarrow
\omega +i0^+$), have no imaginary part, and hence the real frequency
response will be dominated by the singularities of the diagonal
Green's function only.

Analytically continuing to real frequencies and working out the
Matsubara sums (and noting that matrix elements of $M$ are real in
this limit), we get
\begin{widetext}
\beq n^{(2)}(\sigma,k,
  \omega)=2\lambda\sum_{pq}\int_{-\infty}^\infty
  \frac{d\omega'}{\pi} \alpha^{\sigma}_{pq}(k,\omega',\omega+\omega') G_D^{p''}(k,\omega')G_D^{q''}(k,\omega+\omega')[n_B(\omega')-n_B(\omega'+\omega)]
\eeq
%
where $p, q =\pm$, $G_D^{p''}$ indicates the imaginary part of the
Green's function component, and the matrix element
%
\bqa
 \alpha^{\sigma}_{pq}(k,\omega_1,\omega_2)&= &\sum_{mn} M_{\sigma
  p}(k,\omega_1)M^{-1}_{pm}(k,\omega_1)\hat{\Lambda}_{mn}(k) M_{n q}(k,\omega_2)M^{-1}_{q\sigma}(k,\omega_2)
\eqa
%
We now restrict ourselves to the response at $T=0$. From the simple
pole structure of the Green's function the response is then obtained as
%
\bqa \no n^{(2)}(\sigma,k,
  \omega)&=&2\pi\lambda
\left [ \alpha^{\sigma}_{++}(k,E^+_{pk},E^+_{hk})
    z^+_{pk}z^+_{hk}\delta(\omega
    -E^+_{pk}+E^+_{hk}) +\alpha^{\sigma}_{+-}(k,E^-_{pk},E^+_{hk})
    z^+_{hk}z^-_{pk}\delta(\omega
    -E^-_{pk}+E^+_{hk})\right.\\
& & \left. +\alpha^{\sigma}_{++}(k,E^+_{hk},E^+_{pk})
    z^+_{pk}z^+_{hk}\delta(\omega
    -E^+_{hk}+E^+_{pk}) +\alpha^{\sigma}_{+-}(k,E^+_{hk},E^-_{pk})
    z^+_{hk}z^-_{pk}\delta(\omega
    -E^+_{hk}+E^-_{pk})\right]
\label{eq:ndeltafn}
\eqa
\end{widetext}
It is clear from Eq.~\ref{eq:ndeltafn} that the response shows up on two
contours specified by $\omega=E^{+}_{pk}-E^+_{hk}$ and
$\omega=E^{-}_{pk}-E^+_{hk}$, the intra and interband particle-hole
excitations. In the Mott phase, both these excitations are gapped with
the lowest Mott gap corresponding to the intra-band excitations. As
the frequency of modulation is swept, there is no response till the
frequency crosses the Mott gap. Beyond this point, two distinctive set
of phenomena can be seen depending on the ratio, $\gamma/J$. For small
$\gamma/J$ the minimum of the intraband excitations is at the zone
center, $[0,0]$. This is seen in Fig.~\ref{fig1}(a), where the
intraband excitation in the Brillouin zone is plotted as a color plot
for the following set of parameters: $J=0.02U$, $\gamma=0.01U$,
$\Omega=0.01 U$, $\mu=0.2U$ and $\zeta=0.4$, where the minimum of the
spectrum (the Mott gap) is
$0.24U$ at the zone center. The dispersion increases as one moves away
from the zone center and the highest excitation energy occur at
$[\pi,\pi$ with $\omega=0.57 U$. The corresponding inter-band
excitations are shown in Fig.~\ref{fig1}(b). They follow a similar
pattern as the intra-band excitations with a minimum energy of $0.73
U$ and a maximum energy of $1.22 U$.
\begin{figure*}[t]
\includegraphics*[width=0.25 \linewidth]{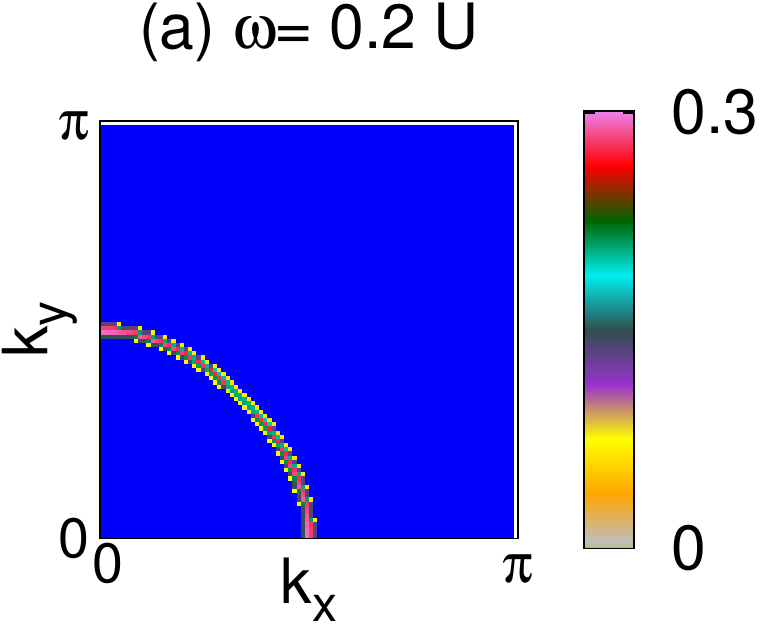}
\includegraphics*[width=0.25 \linewidth]{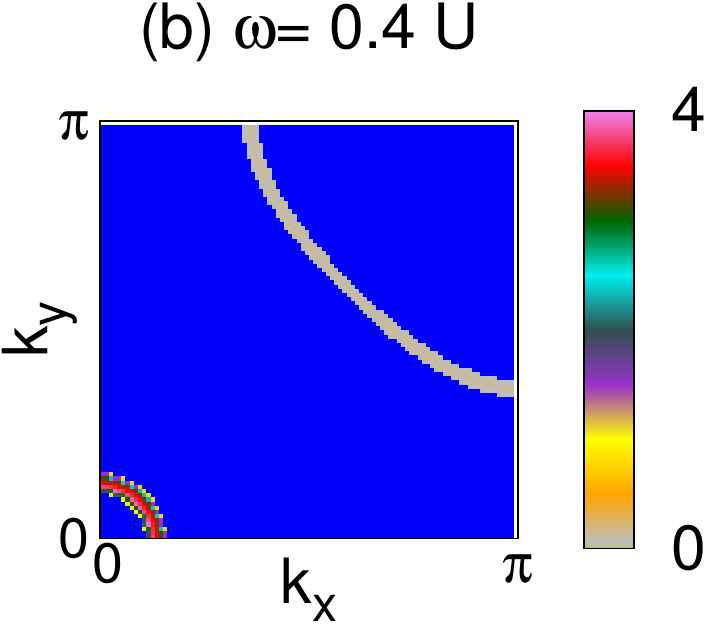}
\includegraphics*[width=0.25 \linewidth]{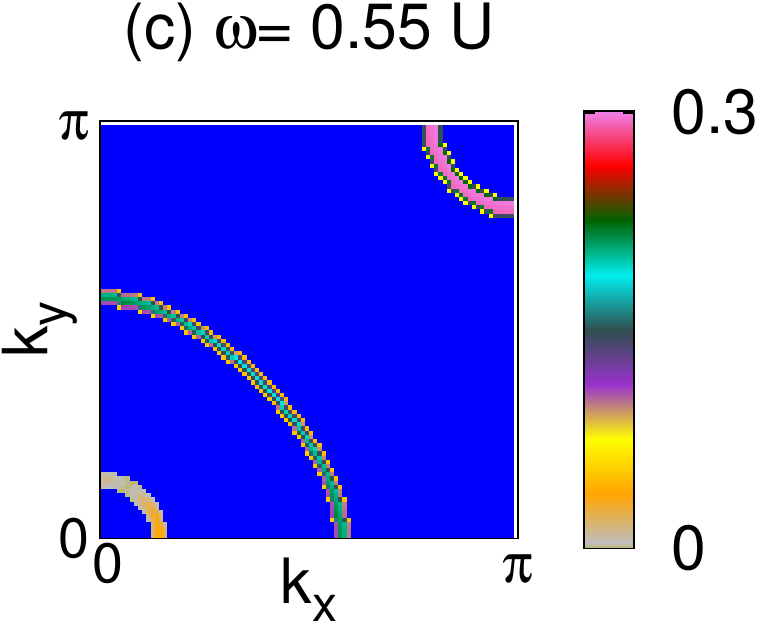}
\includegraphics*[width=0.25 \linewidth]{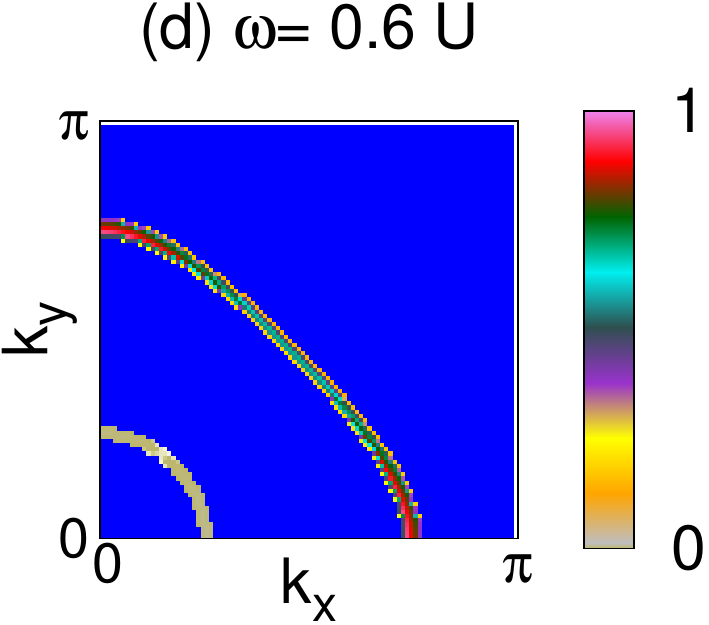}
\includegraphics*[width=0.25 \linewidth]{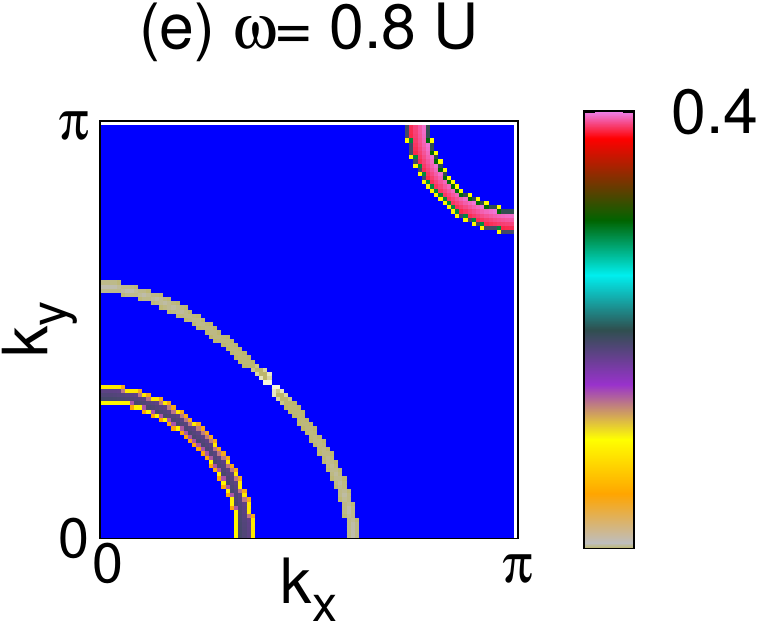}
\includegraphics*[width=0.25 \linewidth]{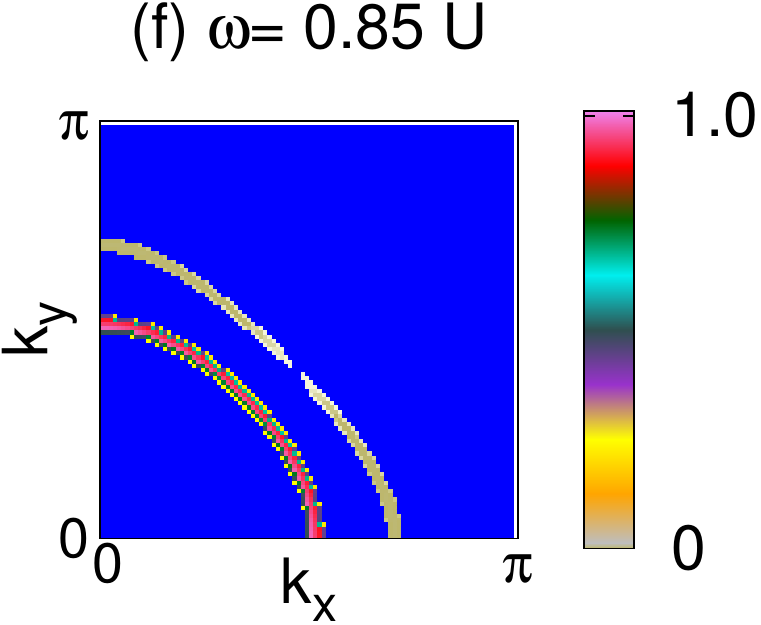}
\includegraphics*[width=0.25 \linewidth]{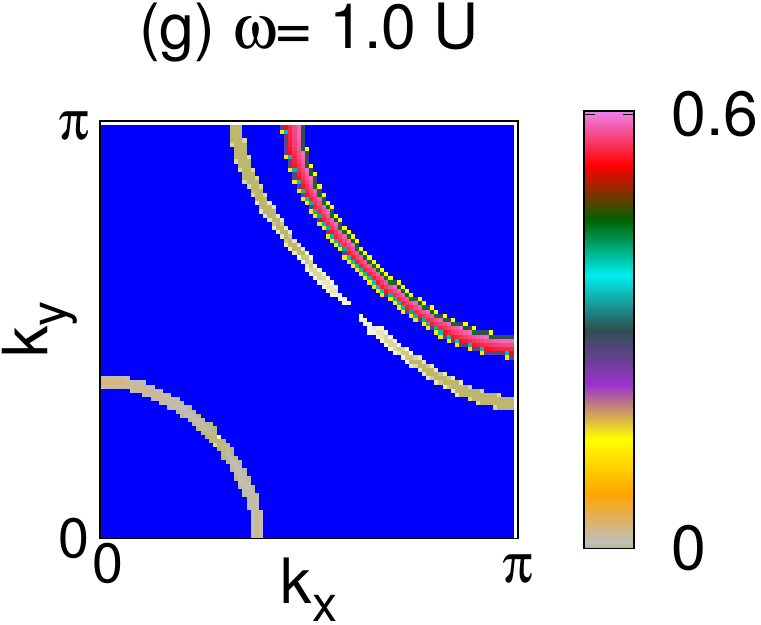}
\includegraphics*[width=0.25 \linewidth]{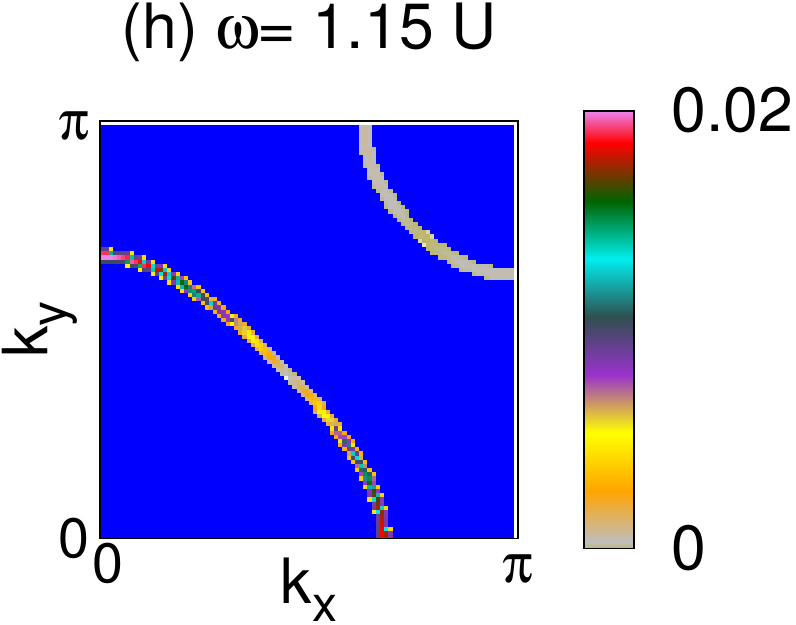}
\includegraphics*[width=0.25 \linewidth]{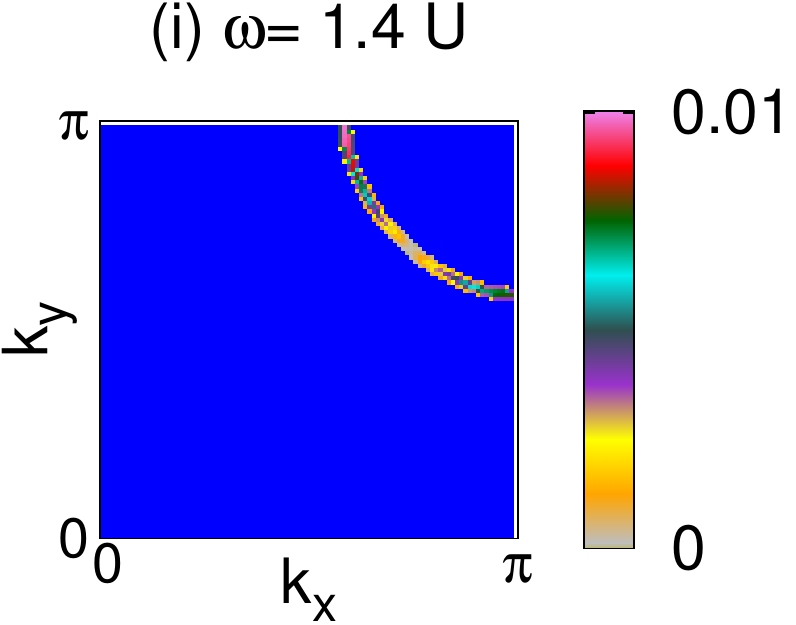}
\caption{Lattice modulation response in the superfluid phase. The
parameters of the system are $J=0.03U$, $\gamma=0.01U$,
$\mu=0.35U$,$\Omega=0.01U$ and $\zeta=0.4$. In this
  case the BEC is formed at the zone center. (a) A
    single contour corresponding to the two phase modes (b) 2
    contours (outward from zone center): a phase and an amplitude
    mode, two phase modes (c) 3 contours  (outward from zone
    center): a phase mode and an interband transition, a phase and
    an amplitude mode, two phase modes (d) 2
    contours (outward from zone center): a phase mode and an
    interband transition, a phase and an amplitude
    mode (e) 3 contours  (outward from zone
    center): two amplitude modes, a phase mode and an interband transition, a phase and
    an amplitude mode (f) 2 contours  (outward from zone
    center):  two amplitude modes, a phase mode and an interband transition (g) 3 contours  (outward from zone
    center): an amplitude mode and an interband transition, two
    amplitude modes, a phase and an interband transition (h) 2 contours  (outward from zone
    center): an amplitude mode and an interband transition, two
    amplitude modes (i) a single contour: an amplitude mode and an interband transition. Note that the spectral weight of the interband transitions are much smaller than the weights of other transitions.}
\label{fig6}
\end{figure*}

The optical modulation response in this case is plotted for four
different frequencies in this case in Fig.~\ref{fig2}. As frequency
increases beyond the gap, $0.24U$, a contour of excitations is seen,
which become larger and moves towards the edge of the Brillouin
zone, before disappearing at $\omega=0.57U$. This contour,
corresponding to intra-band excitations, are shown in
Fig~\ref{fig2}(a) and (b). As the frequency is further increased
beyond $\omega=0.73U$, the contours corresponding to the inter-band
excitations appear around the zone center and move outwards, before
disappearing at $\omega=1.22U$, as shown in Fig~\ref{fig2}(c) and
(d). We would like to note here that the full dispersion of the
excitations can be tracked from the optical lattice modulation
spectroscopy. To show how this is done, in Fig.~\ref{fig8} we plot
the lattice modulation response in the Mott phase as a function of
frequency along various cuts in the Brillouin zone, going radially
outwards at different angles $\theta$ with the $k_x$ axis. Fig.
~ref{fig8} (a)-(c) correspond to the system parameters $J=0.02U$,
  $\gamma=0.01U$, $\mu=0.2U$,$\Omega=0.01U$ and $\zeta=0.4$ with the
  band minimum at the zone center, which is clearly seen from the plots.
Further, the modulation spectroscopy also provide information about
the spectral weight of the various excitations. In this case, it is
evident the interband excitations carry much less spectral weight
than the intraband excitations.

The phenomenology changes dramatically if the spin orbit coupling
$\gamma$ is larger than hopping $J$. Fig.~\ref{fig1}(c) and (d)
shows a plot of the intra and interband particle hole excitations in
the Mott phase for a system with parameter values $J=0.01U$,
$\gamma=0.04U$, $\Omega=0.01 U$, $\mu=0.2U$ and $\zeta=0.4$. In this
case the spin orbit coupling is larger than the hopping and
consequently the intraband excitations have a minimum at $[\pm k_0,
\pm k_0]$ along the zone diagonal. The location of the minima
corresponding $[k_0,k_0]$ is shown in Fig.~\ref{fig1}(c); the
corresponding figures for other minima can be obtained by a
reflection of the contours in the first quadrant ($0\le k_x,k_y\le
\pi)$ about the $k_x$ and $k_y$ axes and the origin.
The interband excitations however have a minimum at the zone center,
as seen in Fig.~\ref{fig1}(d). Once again, the system will not show
any response as long as the modulation frequency is below the Mott
gap of $0.28U$. Once the Mott gap is crossed, excitation contours
around $[\pm k_0,\pm k_0]$ appear in the response, as seen (for
$[k_0,k_0]$) in Fig~\ref{fig3}(a). These contours spread out and
disappear at $\omega=0.45U$, as seen in Fig~\ref{fig3}(b). Another
contour, corresponding to interband excitations and centered around
$k=0$ appears as the modulation frequency is swept beyond $0.87U$.
This spreads out and finally disappears at $\omega=1.1U$, the upper
limit of the interband excitation energy, as seen in
Fig.~\ref{fig3}(c) and (d). To get a better idea of the dispersion,
the lattice modulation response for a system in Mott phase with a
band minimum shifted to finite wavevectors is plotted in Fig.
~\ref{fig8} (d) - (f), where the parameters of the system are
$J=0.01U$,
  $\gamma=0.04U$, $\mu=0.2U$,$\Omega=0.01U$ and $\zeta=0.4$. In this case, the
  band minimum is at $(k_0,k_0)$ with $k_0=1.131 a^{-1}$. It is
  clearly seen that the intraband excitations have a shifted band
  minimum, while the interband excitations continue to have a minima
  at the zone center. In this case, it is clear from comparison
with corresponding response in Fig.~\ref{fig2}, that although the
intraband excitations have larger spectral weight than the interband
excitations, the interband excitations carry relatively larger
spectral weight than the case where the minimum was at the zone
center.

The lattice modulation spectroscopy can identify the Mott phase from
the existence of a gap in the spectrum. However the momentum
resolved nature of the spectroscopy provides much more detailed
information about the single particle spectral function of the
system. It provides information about the spectrum and one should be
clearly able to see the minimum of the excitation spectrum shift
from the zone center as the parameters are varied. The modulation
spectroscopy also provides information about the spectral weight of
the excitations of the system, which is of immediate relevance for
figuring out both near equilibrium and far from equilibrium response
of the system to different stimuli.

\section {Response in the Superfluid phase \label{optmod_sf}}
\begin{figure}[t!]
\includegraphics*[width=0.47 \linewidth]{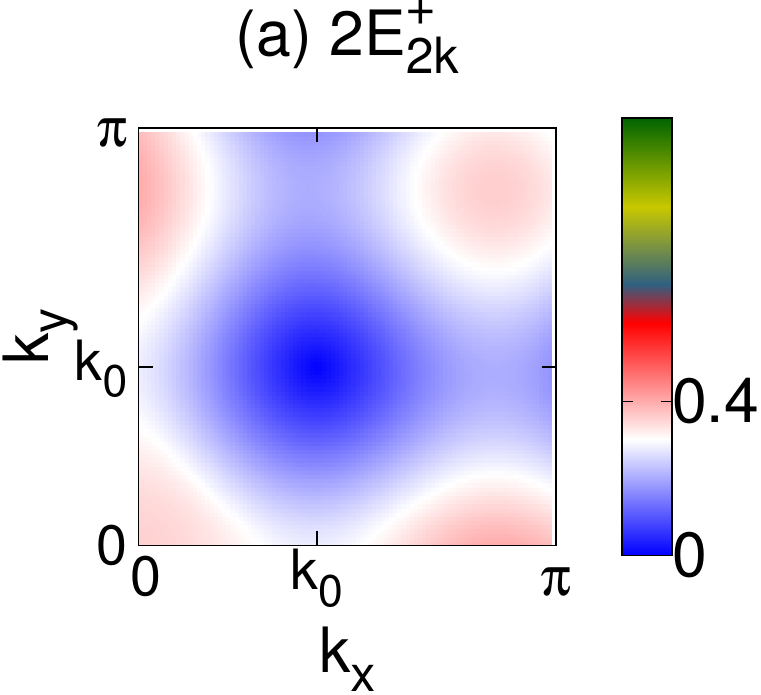}
\includegraphics*[width=0.47 \linewidth]{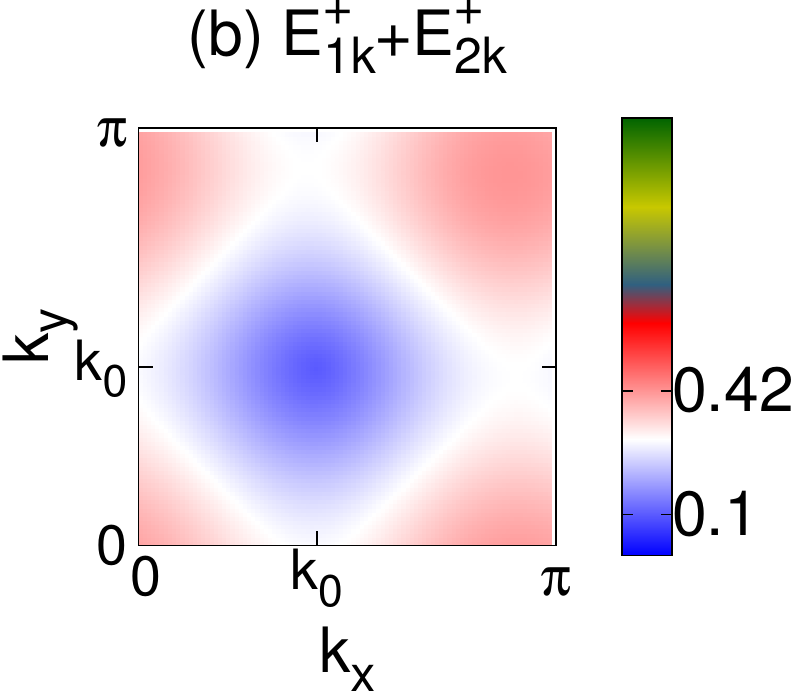}
\includegraphics*[width=0.47 \linewidth]{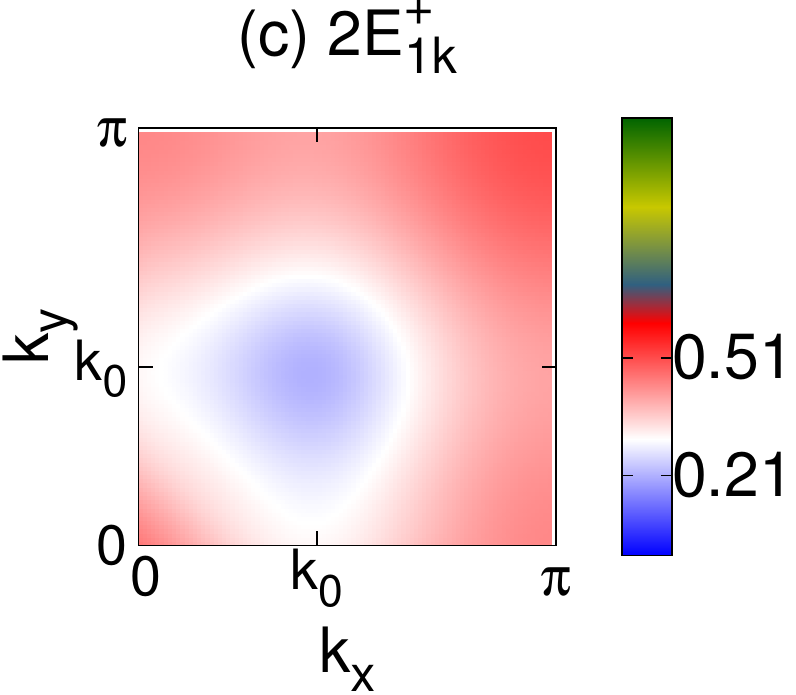}
\includegraphics*[width=0.47 \linewidth]{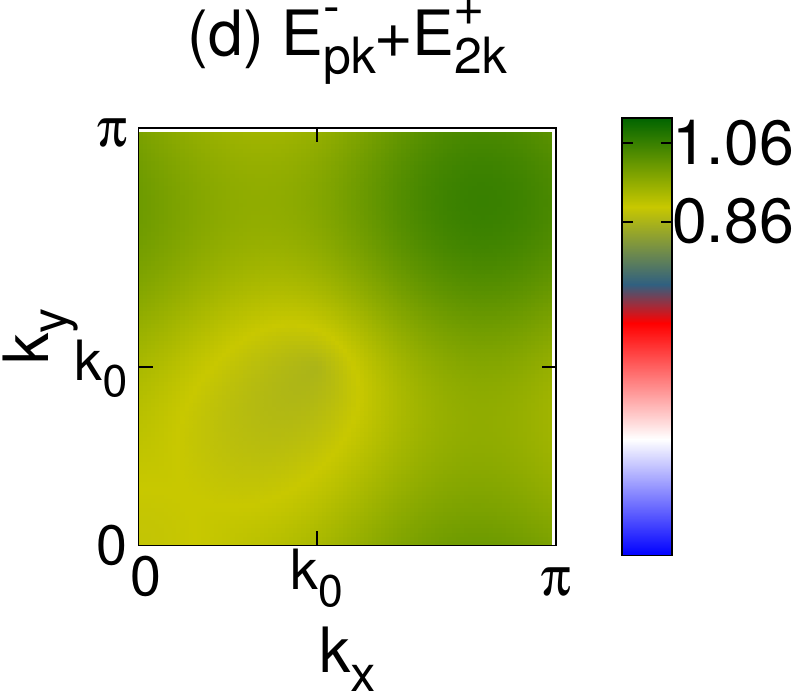}
\includegraphics*[width=0.47 \linewidth]{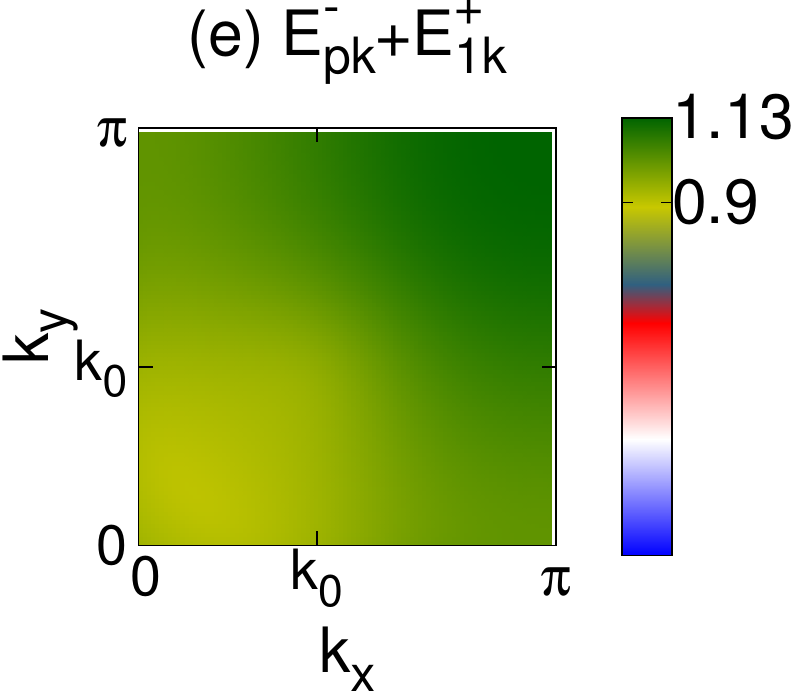}
\caption{Excitation spectrum in the superfluid phase. The parameters
of the system are $J=0.01U$,
  $\gamma=0.06U$, $\mu=0.2U$,$\Omega=0.01U$ and $\zeta=0.4$. In this
  case the BEC is formed at a finite momentum $[\pm k_0,\pm k_0]$, with $k_0=1.34
  a^{-1}$, away from the zone center. The excitations are (a) two
phase modes (b) one phase and one amplitude mode (c) one phase mode
and one interband transition (d) two amplitude modes and (e) an
amplitude mode and an interband transition. Note that the color
scheme has same absolute value in all panels to show which
excitations overlap with each other in energy. } \label{fig5}
\end{figure}

The spin-orbit coupled bosons undergo a Mott insulator-superfluid
quantum phase transition as either the hopping or the spin-orbit
coupling is increased as both terms help to delocalize the bosons.
If the transition occurs at a large value of $J/\gamma$, the
superfluid phase has a BEC at the zone center $k=[0,0]$ with
associated anomalous propagators and Goldstone modes. On the other
hand, if the transition takes place at a large value of $\gamma/J$,
the system exhibits the twisted superfluid phase, with a BEC at a
finite momentum. In general there can be four such momentum values
given by $[\pm k_0,\pm k_0]$ which are degenerate minima of the
spectrum. This in principle allows the possibility of formation of a
square superlattice (i.e. a supersolid) with the incommensurate
momentum $k_0$ providing the inverse lattice constant. However, in a
cold atom system with the presence of a trapping potential which
breaks translational symmetry, the more likely effect is the
formation of domains~\cite{powell}, each of which corresponds to
choosing one of the four values of allowed $k_0$. Assuming domains
of size much larger than $k_0^{-1}$, the momentum distribution
signal will be an incoherent weighted sum of the signal from a
single domain with a fixed condensation wavevector. In this paper,
we shall assume that condensation occurs at only one possible
momentum which we choose to be $(k_0,k_0)$ and work out the signal
from a single domain. The important features, like the low energy
features around$(k_0,k_0)$ will occur in different parts of the
Brillouin zone for different domains and thus will not be washed out
by formation of domains. Although the detailed signal requires
knowledge of domain distribution, a first approximation can be
obtained from our calculation by symmetry, together with an idea of
the density of each type of domains. The lattice modulation
spectroscopy is uniquely suited to distinguish between gapless
superfluid phase from the Mott phase, and owing to its momentum
resolved nature, it can also distinguish between the normal and
twisted superfluid phases. Since this method gives detailed
information about the spectral function, useful quantities like the
speed of sound can be calculated from the spectrum, while the
relative intensity of the contours will provide information about
transfer of spectral weight among the different kinds of excitations
as the energy is varied.

In the superfluid phase the bosons occupying the lower energy band
form a BEC at the appropriate momentum. In this case the Green's
functions expand to a $4\times 4$ matrix to accommodate the
anomalous propagators. It is easier to work in the $\pm$ basis,
earlier described in the Mott phase, since the structure of the
Green's functions are simplest in this basis. Working with a $4$
component vector
$[\phi^-(k,i\omega),\phi^{-\ast}(2k_0-k,-i\omega),\phi^+(k,i\omega),\phi^{+\ast}(2k_0-k,-i\omega)]$,
the inverse Green's function is given by
\cite{mandal,sensarma1,sengupta1}
\begin{widetext}
\beq
G^{-1}(k,i\omega_n)\simeq\left(\begin{array}{cccc}
  \zeta_-(k,i\omega_n)-r  & 0 &0 &0\\
0 &  \zeta_-(2k_0-k,-i\omega_n)-r  & 0 &0\\
0 & 0 &\zeta_+(k,i\omega_n)-r  & -r\\
0 & 0 & -r &\zeta_+(2k_0-k,-i\omega_n)-r  \\
\end{array}\right)
\eeq
\end{widetext}
where $\zeta_{\pm}(k,i\omega)$ is defined by Eq.~\ref{eq:zeta} and $r=\zeta_+(k_0,k_0,\omega=0)$ incorporates
the effects of the
presence of the condensate. The lower $2 \times 2$ block simply looks
like the Green's function of a Bose gas in the Bogoliubov
approximation, with the complicated function $\zeta_+(k,\omega)$,
incorporating correlations due to proximity to a Mott insulator,
replacing a simple free particle piece. In the diagonal upper block,
the presence of the condensate leads to a Hartree correction due to
inter-band interactions.
\begin{figure*}[t]
\includegraphics[width=0.24 \linewidth]{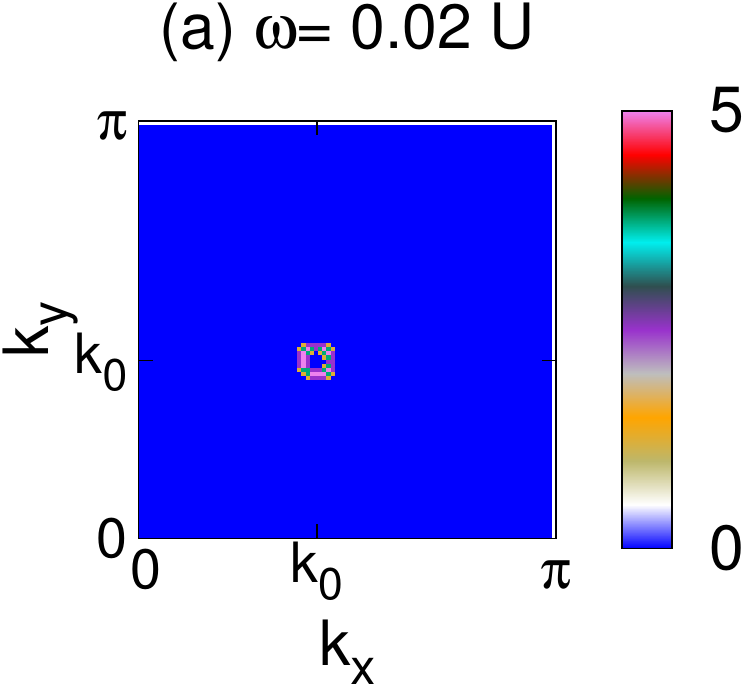}
\includegraphics[width=0.24 \linewidth]{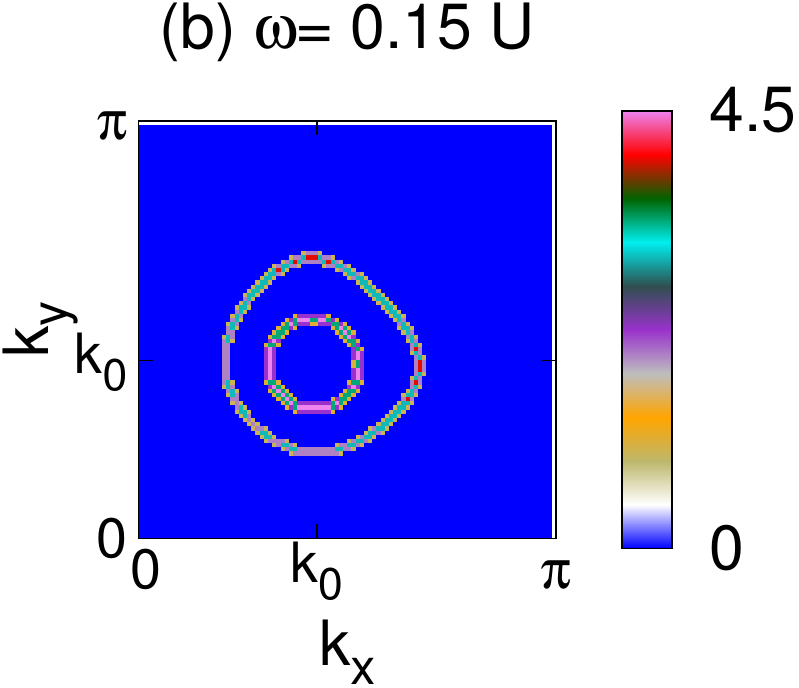}
\includegraphics[width=0.24 \linewidth]{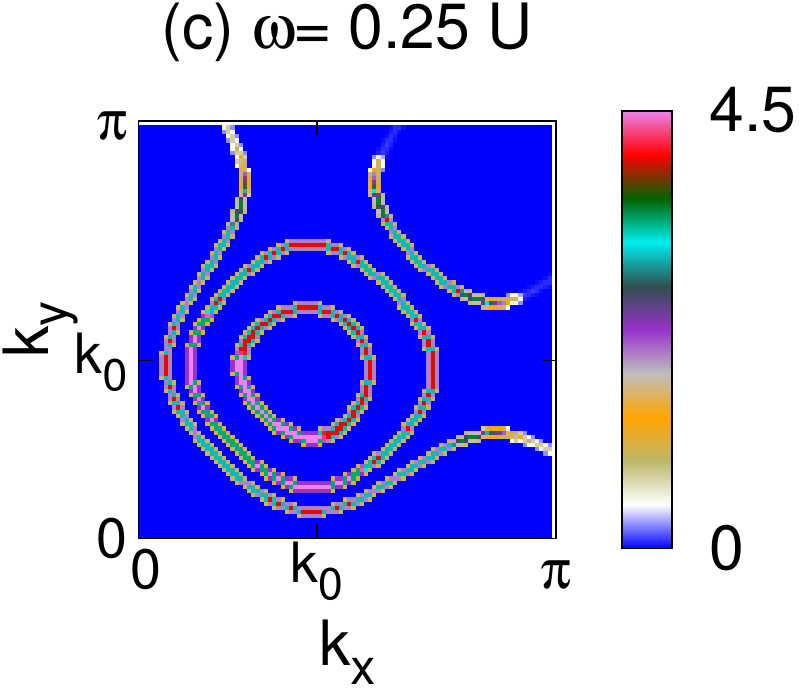}
\includegraphics[width=0.24 \linewidth]{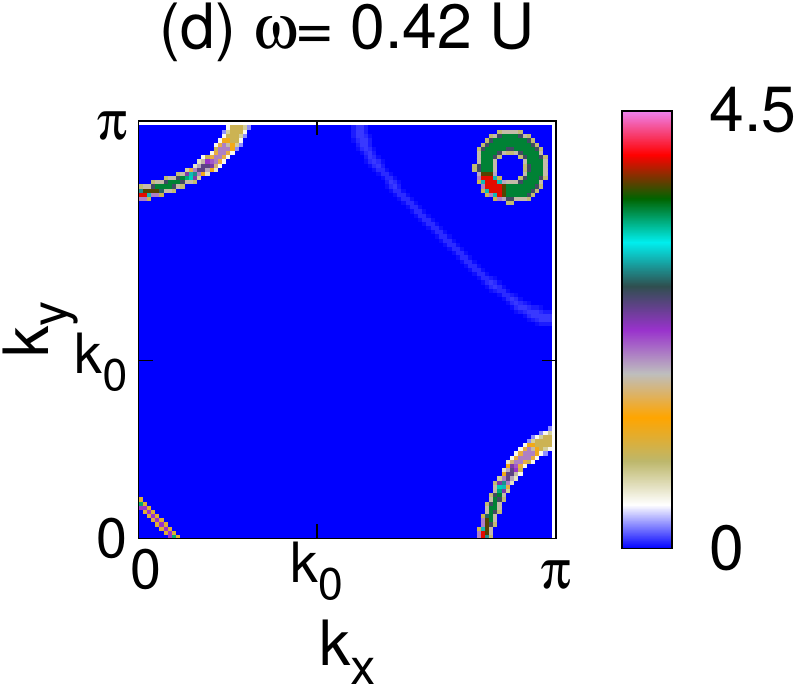}
\includegraphics[width=0.24 \linewidth]{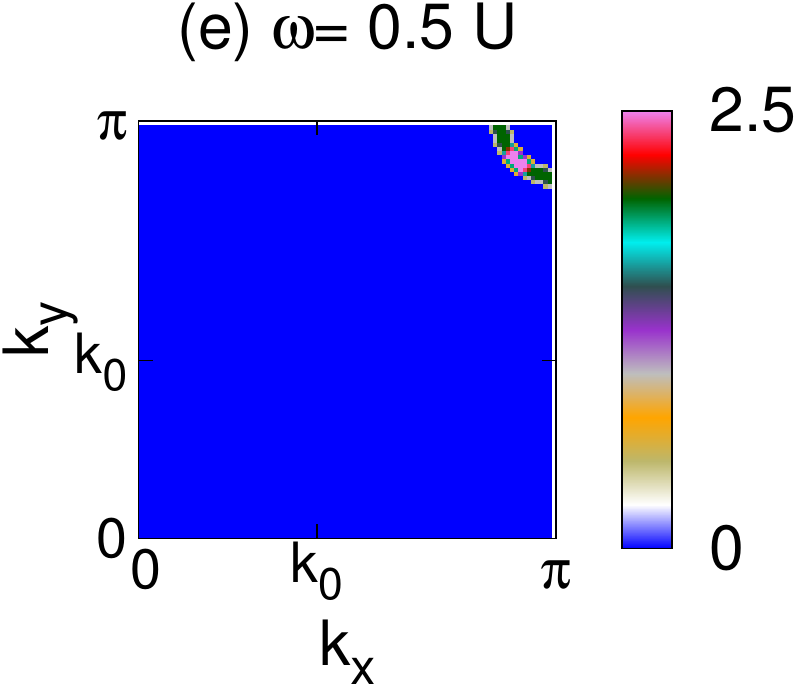}
\includegraphics[width=0.24 \linewidth]{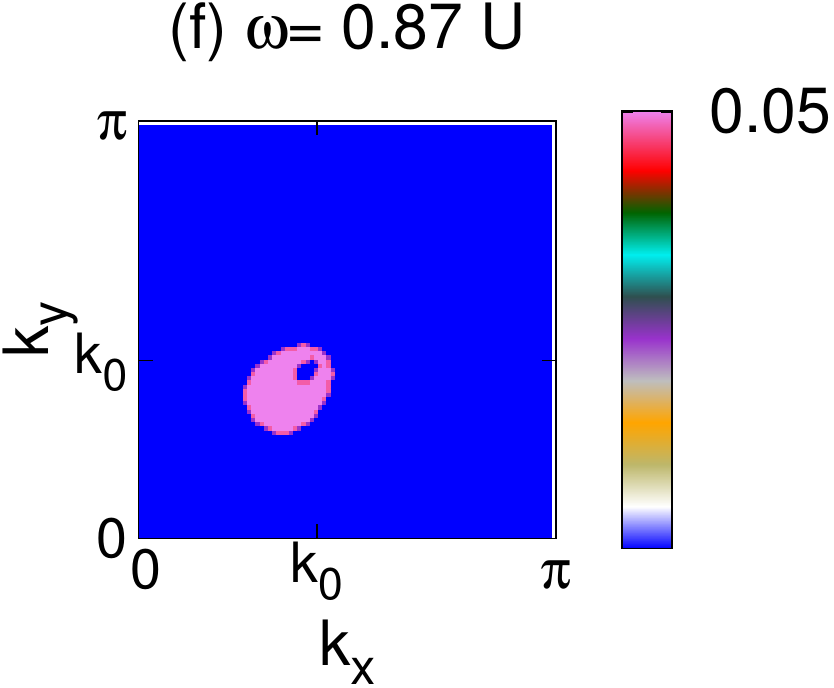}
\includegraphics[width=0.24 \linewidth]{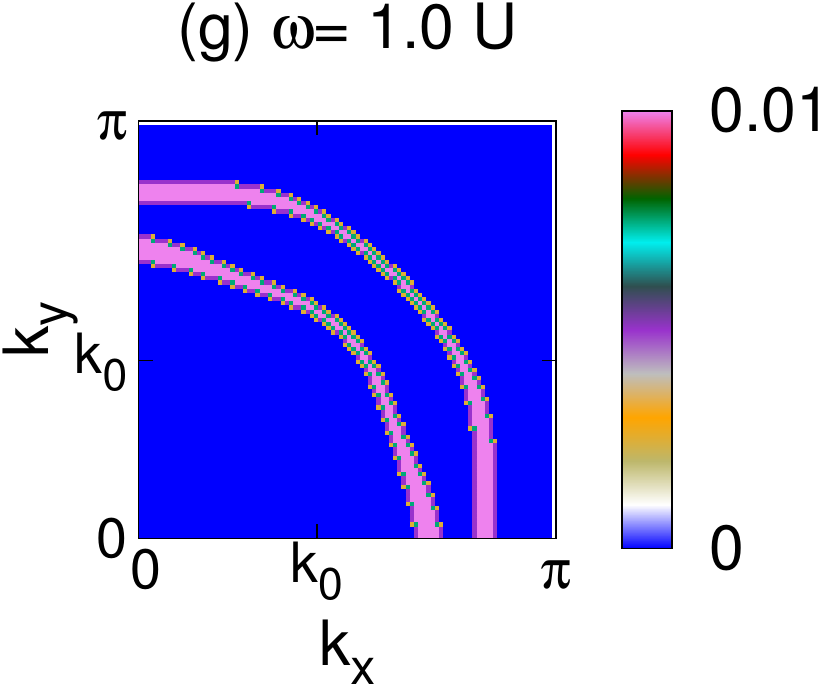}
\includegraphics[width=0.24 \linewidth]{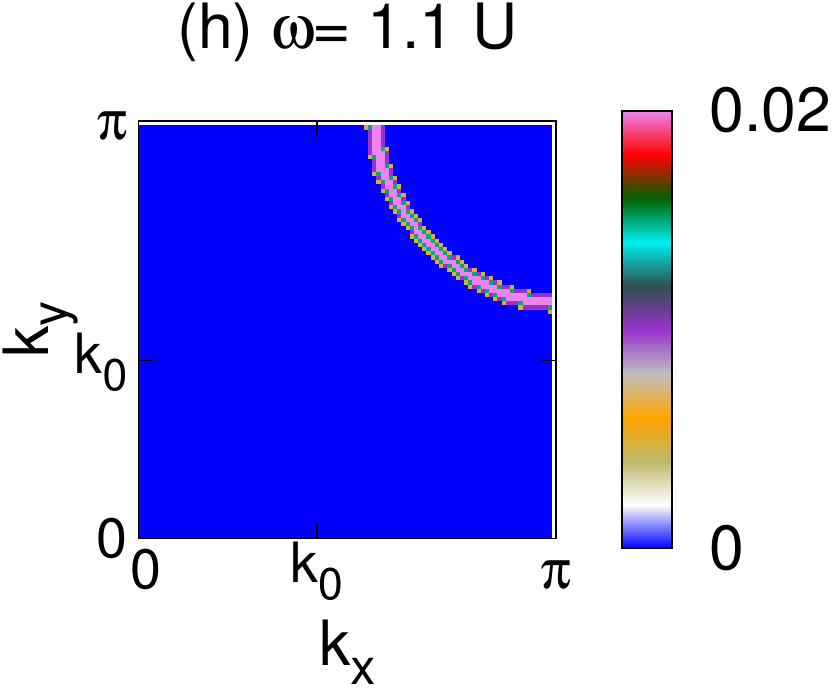}
\caption{Lattice modulation response in the superfluid phase. The
parameters of the system are $J=0.01U$,
  $\gamma=0.06U$, $\mu=0.2U$,$\Omega=0.01U$ and $\zeta=0.4$. In this
  case the BEC is formed at a finite momentum $[k_0,k_0]$ ($k_0=1.34
  a^{-1}$) away from the zone center. A
    single contour corresponding to the two phase modes (b) 2
    contours (outward from zone center): a phase and an amplitude
    mode, two phase modes (c) 4 contours  : two open contours due to
    two phase modes, and two closed contours (outward from the zone center) due to
    two amplitude modes, a phase and
    an amplitude mode (d) closed contour due to a phase and an amplitude
    mode and open contours due to two amplitude modes,  (e) a single contour: two amplitude modes
    (f) a single contour: a phase mode and an interband transition at finite momentum (g) 2 contours
    (outward from zone center): an amplitude mode and an interband transition,  a phase and an
    interband transition (h) a single contour: an amplitude mode and an interband transition. Note that the spectral
weight of the interband transitions are much smaller than the
weights of other transitions. } \label{fig7}
\end{figure*}

The unitary transform which converts the original ``spin'' basis,
i.e. $[\phi_\up(k,i\omega),\phi^{\ast}_\up(2k_0-k,-i\omega),\phi_\dn(k,i\omega),\phi^{\ast}_\dn(2k_0-k,-i\omega)]$,
to the $\pm$ basis is given by
\begin{widetext}
\beq
\tilde{M}(k,i\omega_n)\simeq\left(\begin{array}{cccc}
  M_{11}^\ast(k,i\omega)  & 0 &M_{21}^\ast(k,i\omega) &0\\
0 &  M_{11}(2k_0-k,-i\omega)  & 0 &M_{21}(2k_0-k,-i\omega)\\
 M_{12}^\ast(k,i\omega)  & 0 &M_{22}^\ast(k,i\omega) &0\\
0 &  M_{12}(2k_0-k,-i\omega)  & 0 &M_{22}(2k_0-k,-i\omega)
\end{array}\right),
\label{eq:sogapp_1}
\eeq
%
 where the matrix elements $M_{ij}$ are given by Eq.\
\ref{eq:somat}. The final Green's function has the form
%
\beq
G(k,i\omega_n)=\left(\begin{array}{cccc}
 G^-(k,i\omega_n)  & 0 &0 &0\\
0 &  G^-(2k_0-k,-i\omega_n)  & 0 &0\\
0 & 0 &G^+(k,i\omega_n;k_0)  & F(k,i\omega_n;k_0)\\
0 & 0 & F(2k_0-k,-i\omega_n;k_0) &G^+(2k_0-k,-i\omega_n;k_0)  \\
\end{array}\right). \label{grdef}
\eeq
\end{widetext}
The full expression for the Green's functions are complicated, but, as
in the case of the Mott phase, the only singularities of the Green's
functions are simple poles. So, as far as the imaginary part of the
Green's function is concerned, they are faithfully reproduced by
\begin{widetext}
\bqa G^-(k,i\omega)&=&\frac{g^-_{pk}}{i\omega -E^-_{pk}},\quad
G^+(k,i\omega;k_0)= \frac{g^{+p}_{1k}}{i\omega
  -E^+_{1k}}+\frac{g^{+h}_{1k}}{i\omega
  +E^+_{1k}}+\frac{g^{+p}_{2k}}{i\omega
  -E^+_{2k}}+\frac{g^{+h}_{2k}}{i\omega +E^+_{2k}}\\
F(k,i\omega;k_0)&=&\frac{f^{+}_{1k}}{i\omega
-E^+_{1k}}-\frac{f^{+}_{1k}}{i\omega
+E^+_{1k}}+\frac{f^{+}_{2k}}{i\omega
-E^+_{2k}}-\frac{f^{+}_{2k}}{i\omega +E^+_{2k}} \eqa
\end{widetext}
\begin{figure*}[t]
\includegraphics[width=0.33 \linewidth]{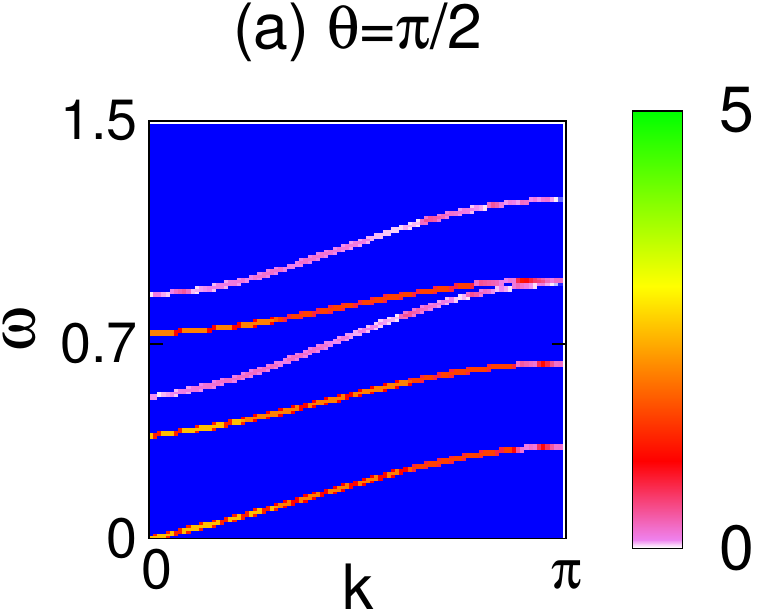}
\includegraphics[width=0.33 \linewidth]{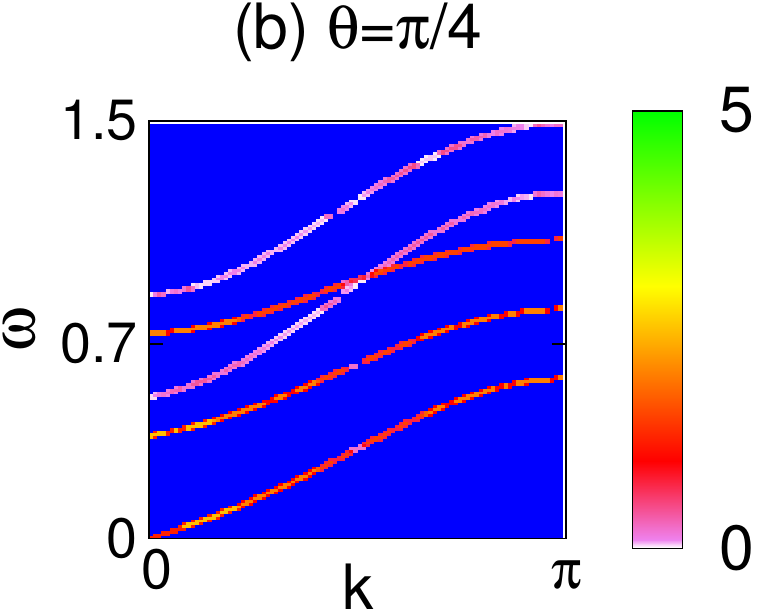}
\includegraphics[width=0.33 \linewidth]{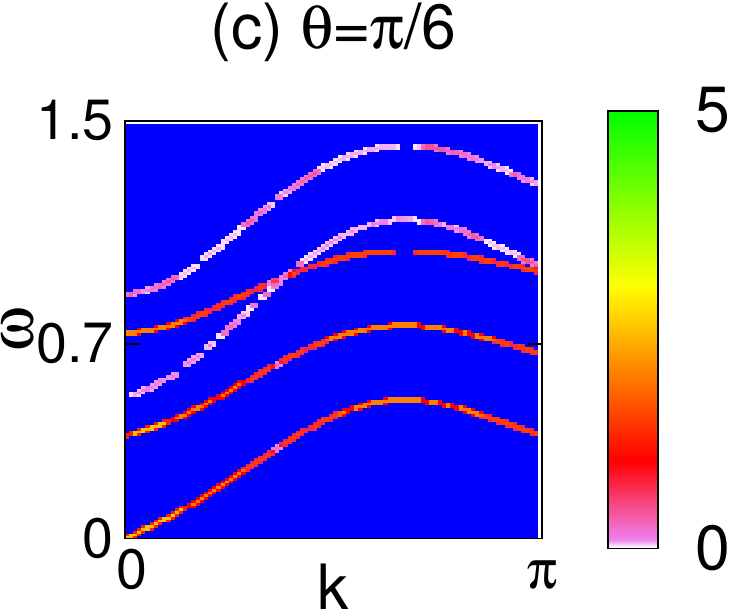}
\includegraphics[width=0.33 \linewidth]{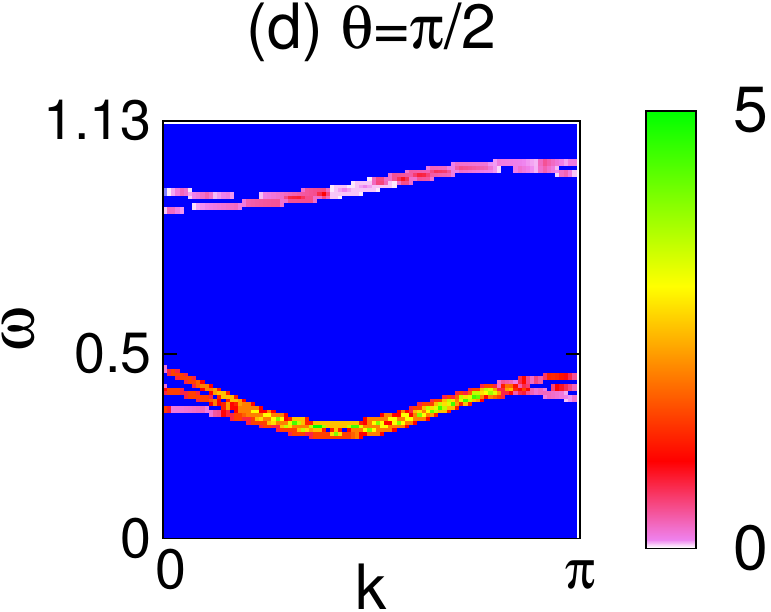}
\includegraphics[width=0.33 \linewidth]{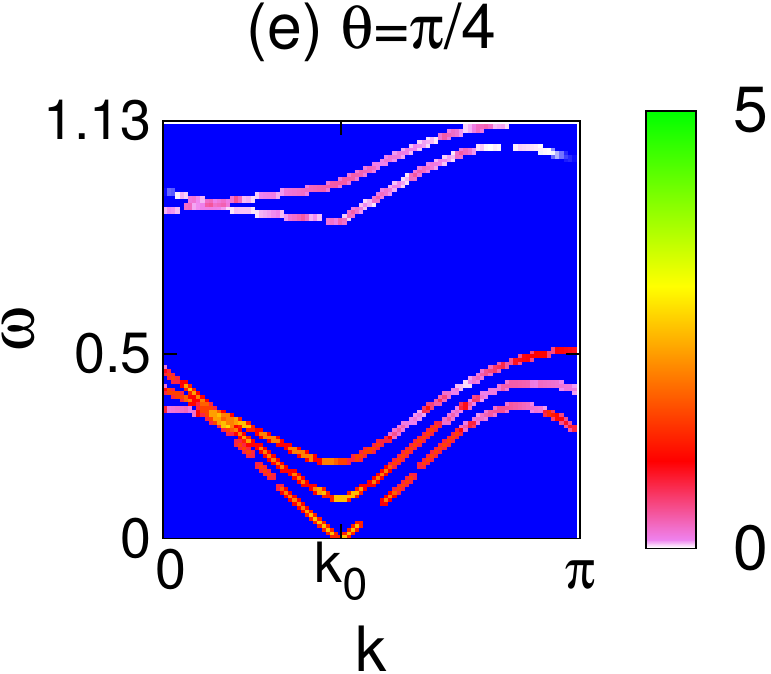}
\includegraphics[width=0.33 \linewidth]{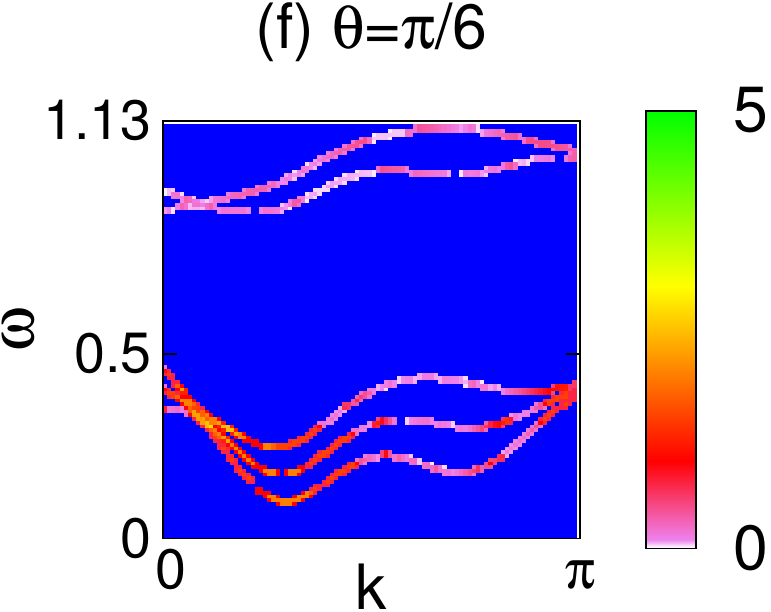}
\caption{Lattice modulation response in the superfluid phase with the
  variation in frequency for various cuts in the Brillouin zone going radially outwards at an angle
$\theta$ with the $k_x$ axis. In (a) - (c) the parameters of the system are $J=0.03U$, $\gamma=0.01U$,
$\mu=0.35U$,$\Omega=0.01U$ and $\zeta=0.4$ with the
  condensate at the zone center. (a) $\theta=\pi/2$, (b)
  $\theta=\pi/4$ and (c) $\theta=\pi/6$. All the five branches
  including the gapless linear Goldstone mode near $k=0$ is clearly seen. In (d) - (f) the parameters
  of the system are  $J=0.01U$, $\gamma=0.06U$, $\mu=0.2U$,$\Omega=0.01U$ and $\zeta=0.4$. In this
  case the condensate is located at $(k_0,k_0)$ with $k_0=1.131 a^{-1}$.
  (d) $\theta=\pi/2$, (e) $\theta=\pi/4$ and (f) $\theta=\pi/6$. Only the cut along $[1,1]$ direction passes through the
condensate location and shows the gapless Goldstone mode, while the
spectrum is gapped along the other cuts. } \label{fig9}
\end{figure*}
Here the particle excitation in the $-$ branch, $E^-_{pk}$ is the
solution of $\zeta_-(k,E^-_{pk})=r$ and the four excitation poles
$\pm E^{+}_{1(2) k}$ are obtained from the solutions of
\begin{eqnarray}
[\zeta_+(k,\omega)-2r][\zeta_+(2k_0-k,-\omega)-2r]-r^2=0.
\label{pluseq}
\end{eqnarray}
The quasiparticle residues are obtained from the frequency
derivatives of the Green's functions at the corresponding poles.
Note that in the $+$ band, the particle and hole excitations are now
mixed due to formation of a condensate and that the energies
$E_{1(2) k}^{+}$, $g^{+p(h)}_{1(2)k}$, and $f^{+}_{1(2)k}$ depends
on $k_0$ through Eq.\ \ref{pluseq}; this is in contrast to the
corresponding quantities in the $-$ band which has no particle-hole
mixing. This necessitates the use of additional $k_0$ argument in
the definition of $G^+$ and $F$ in Eq.\ \ref{grdef}; however, we
refrain from putting such additional label of $k_0$ in $E_{1(2)
k}^{+}$, $g^{+p(h)}_{1(2)k}$, and $f^{+}_{1(2)k}$ for notational
brevity.

The response function for the lattice modulation spectroscopy is then
given by
\begin{widetext}
\bqa \no \Pi_\sigma(k, i\omega_n)&=&\frac{\lambda}{\beta}\sum_{
i\omega_l}\left[
\tilde{M}(k,i\omega_l)G(k,i\omega_l)\tilde{M}^{-1}(k,i\omega_l)
\hat{\Lambda}^{'}(k)\tilde{M}(k,i\omega_l+i\omega_n)G(k,i\omega_l+i\omega_n)\tilde{M}^{-1}(k,i\omega_l+i\omega_n)\right]_{2a-1,2a-1}\\
& &+\omega_n \rightarrow -\omega_n
\eqa
%
where $a=1$ for $\sigma=\up$ and $a=2$ for $\sigma=\dn$, and the
perturbation matrix $\Lambda'$ is given by
\beq
\Lambda^{'}(k)=\frac{1}{2}\left(\begin{array}{cccc}
 \epsilon_k  & 0 &\gamma_k &0\\
0 &  \epsilon_k  & 0 &\gamma^\ast_{-k}\\
\gamma^\ast_k & 0 &\epsilon_k  & 0\\
0 & \gamma_{-k} & 0 &\epsilon_k
\end{array}\right)
\eeq
 Working out
the Matsubara sums, and analytically continuing to real frequencies,
\bqa \no n^{(2)}(\sigma,k,
  \omega)&=&2\lambda\left[\sum_{pq}\int_{-\infty}^\infty
  \frac{d\omega'}{\pi} \alpha^{\sigma}_{pq}(k,\omega',\omega+\omega')
  G^{p''}(k,\omega';k_0)G^{q''}(k,\omega+\omega';k_0)[n_B(\omega')-n_B(\omega'+\omega)]\right.\\
& &\left. +\beta^{\sigma}(k,\omega',\omega+\omega')
F^{''}(k,\omega';k_0)F^{''}(k,\omega-\omega';k_0)[n_B(\omega')-n_B(\omega-\omega')\right]
\eqa
%
where $p, q =\pm$, $G^{p''}$ and $F^{''}$ indicate the imaginary
part of the Green's function component, and it is understood that
$G^{-''}(k,\omega;k_0) \equiv G^{-''}(k,\omega)$. Here the matrix
elements
%
\bqa
 \alpha^{\sigma}_{pq}(k,\omega_1,\omega_2)&= &\sum_{mn} \tilde{M}_{2\sigma-1
  p}(k,\omega_1)\tilde{M}^{-1}_{pm}(k,\omega_1)\hat{\Lambda}^{'}_{mn}(k)\tilde{ M}_{n
  q}(k,\omega_2)\tilde{M}^{-1}_{q 2\sigma-1}(k,\omega_2)\no\\
  \beta^{\sigma}(k,\omega_1,\omega_2)&= &\sum_{mn}  \tilde{M}_{2\sigma-1,3}(k,\omega_1)\tilde{M}^{-1}_{2m,4}(k,\omega_1)
  \hat{\Lambda}^{'}_{2n,2m}(k)\tilde{ M}_{2n,4}(k,\omega_2)\tilde{M}^{-1}_{2\sigma-1,3}(k,\omega_2)
\eqa
%
We now restrict ourselves to the response at $T=0$ for $\omega>0$.
The $\omega <0$ response can then be obtained  from the fact that
the imaginary part of the response function is an odd function of
$\omega$. From the simple pole structure of the Green's function the
response is then obtained as
%
\bqa \no n^{(2)}(\sigma,k,
  \omega)&=&2\pi\lambda
\left [ \rho^{\sigma}_{1}(k)
  \delta(\omega
    -E^-_{pk}-E^+_{1k}) +\rho^{\sigma}_{2}(k)
  \delta(\omega
    -E^-_{pk}-E^+_{2k})\right.\\
& & \left. \rho^{\sigma}_{3}(k)
  \delta(\omega
    -2E^+_{1k}) +\rho^{\sigma}_{4}(k)
  \delta(\omega
    -2E^+_{2k})+\rho^{\sigma}_{5}(k)
  \delta(\omega
    -E^+_{1k}-E^+_{2k}) \right]
\label{eq:ndeltafnsf}
\eqa
%
where the weight of the different contours are given by
\bqa
\no \rho^\sigma_{1}(k)&=&2g^{-}_{pk}g^{+h}_{1k}
\alpha^\sigma_{12}(k,E^{-}_{pk},-E^{+}_{1k}) ~~~~~~~~~~~~~~~~~~~ \rho^\sigma_{2}(k)=2g^{-}_{pk}g^{+h}_{2k}
\alpha^\sigma_{12}(k,E^{-}_{pk},-E^{+}_{2k})\\
 \rho^\sigma_{3}(k)&=&2g^{+p}_{1k}g^{+h}_{1k}
\alpha^\sigma_{22}(k,E^{+}_{1k},-E^{+}_{1k})-(f^{+}_{1k})^2[\beta^\sigma(k,E^{+}_{1k},E^{+}_{1k})+\beta^\sigma(k,-E^{+}_{1k},-E^{+}_{1k})]\\
\no \rho^\sigma_{4}(k)&=&2g^{+p}_{2k}g^{+h}_{2k}
\alpha^\sigma_{22}(k,E^{+}_{2k},-E^{+}_{2k})-(f^{+}_{2k})^2[\beta^\sigma(k,E^{+}_{2k},E^{+}_{2k})+\beta^\sigma(k,-E^{+}_{2k},-E^{+}_{2k})]\\
\no \rho^\sigma_{5}(k)&=&2g^{+p}_{1k}g^{+h}_{2k}
\alpha^\sigma_{22}(k,E^{+}_{1k},-E^{+}_{2k})+2g^{+p}_{2k}g^{+h}_{2k}
\alpha^\sigma_{22}(k,-E^{+}_{1k},E^{+}_{2k})-2f^{+}_{1k}f^{+}_{2k}[\beta^\sigma(k,E^{+}_{1k},E^{+}_{2k})+\beta^\sigma(k,-E^{+}_{2k},-E^{+}_{1k})]
\eqa
\end{widetext}
It is clearly seen that the system would show response on the
contours which correspond to energies $2E^+_{1k}$, $2E^+_{2k}$,
$E^+_{1k}+E^+_{2k}$, $E^-_{pk}+E^+_{1k}$ and $E^-_{pk}+E^+_{2k}$.
Here $E^+_{2k}$ is the Goldstone phase mode which goes gapless,
$E^+_{1k}$ is the gapped amplitude or Higgs mode and $E^-_{pk}$ is
the particle excitation to the upper band, which also has an
excitation gap. The location of these contours in the Brillouin zone
are dramatically different depending on whether the BEC is formed at
the zone center (large $J/\gamma$) or at a finite momentum (small
$J/\gamma$).

The dispersions of  $2E^+_{1k}$, $2E^+_{2k}$, $E^+_{1k}+E^+_{2k}$,
$E^-_{pk}+E^+_{1k}$ and $E^-_{pk}+E^+_{2k}$ are plotted in
Fig.~\ref{fig4} for a system with following set of parameters:
$J=0.03U$, $\gamma=0.01U$, $\Omega=0.01 U$, $\mu=0.35U$ and
$\zeta=0.4$. In this case, the system is in a superfluid with a BEC
at the zone center. Fig.~\ref{fig4} (a) shows $2E^{+}_{2k}$, which
ranges from $0$ at the zone center (the gapless point) to $0.58U$.
All the other plots in Fig.~\ref{fig4} show similar features except
the fact that these excitations are gapped. In Fig.~\ref{fig4} (b),
$E^+_{1k}+E^+_{2k}$  ranges from $0.37 U$ to $0.83U$. Similarly
$E^-_{pk}+E^+_{2k}$ [Fig.~\ref{fig4} (c)]  ranges from $0.51 U$ to
$1.2U$, $2E^+_{1k}$ [Fig.~\ref{fig4} (d)]  ranges from $0.74 U$ to
$1.1U$, and $E^-_{pk}+E^+_{1k}$ [Fig.~\ref{fig4} (e)]  ranges from
$0.88 U$ to $1.5U$.

The optical modulation response for the above system is shown in
Fig~\ref{fig6}. Fig.~\ref{fig6}(a) shows the response at
$\omega=0.2U$, where a single contour of excitations corresponding
to exciting two phase modes, $\omega=2E^{+}_{2k}$ is seen.
Fig.~\ref{fig6}(b) shows the response at $\omega=0.4U$, where two
contours of excitations are seen, the outer one corresponding to
exciting two phase modes, $\omega=2E^{+}_{2k}$, and the inner one
corresponding to exciting a phase and an amplitude mode
$\omega=E^{+}_{2k}+E^{+}_{1k}$.  Fig.~\ref{fig6}(c) shows the
response at $\omega=0.55U$, where three contours of excitations are
seen, the outermost one corresponding to exciting two phase modes,
$\omega=2E^{+}_{2k}$, the middle one corresponding to exciting a
phase and an amplitude mode $\omega=E^{+}_{2k}+E^{+}_{1k}$, and the
innermost contour corresponding to exciting a phase mode and a band
transition to the $-$ band, $\omega=E^{+}_{2k}+E^{-}_{pk}$. As the
frequency is increased to $\omega=0.6U$, the two phase modes
disappear and two contours corresponding to
$\omega=E^{+}_{2k}+E^{+}_{1k}$ and $\omega=E^{+}_{2k}+E^{-}_{pk}$
remain, as seen in Fig. ~\ref{fig6}(d). As the frequency is further
increased to $\omega=0.8U$, an additional contour (the innermost)
corresponding to $\omega=2E^{+}_{1k}$ appears in Fig.~\ref{fig6}(e)
along with the other excitations seen in  Fig. ~\ref{fig6}(d). At
$\omega=0.85U$, the contours in Fig~\ref{fig6}(f) corresponds to
$\omega=2E^{+}_{1k}$ and $\omega=E^{+}_{2k}+E^{-}_{pk}$, while at
$\omega=U$, Fig~\ref{fig6}(g) has an additional innermost contour of
$\omega=E^{+}_{1k}+E^{-}_{pk}$. Finally, in Fig~\ref{fig6}(h), there
are two contours corresponding to $\omega=2E^{+}_{1k}$ and
$\omega=E^{+}_{1k}+E^{-}_{pk}$ at $\omega =1.15U$, while at
$\omega=1.4U$ (Fig.~\ref{fig6}(i) ), only the contour corresponding
to $\omega=E^{+}_{1k}+E^{-}_{pk}$ remains.The presence the gapless
phase mode as well as four other gapped modes is clearly seen in
Fig.~\ref{fig9} (a)-(c), where the lattice modulation response is
plotted as a function of frequency along different radial cuts in
the Brillouin zone making angles $\theta=\pi/2$ , $\theta=\pi/4$ and
$\theta=\pi/6$ with the $k_x$ axis respectively. The speed of sound
in the system can be obtained from the slope of linearly dispersing
phase modes seen in the figure. The gapless mode and the unique
pattern of obtaining up to three contours at certain frequencies
distinguishes the superfluid phase from the Mott phase and can be
used to detect the superfluid-insulator quantum phase transition in
this system. One can also obtain detailed information about the
spectrum and relative spectral weights of the various modes from the
lattice modulation response.

We now consider the changes that appear in the optical modulation
spectroscopy when the spin-orbit coupling is greater than the
hopping amplitude and a BEC is formed at a finite momentum. To see
this, in Fig.~\ref{fig5}, we plot the dispersions of  $2E^+_{1k}$,
$2E^+_{2k}$, $E^+_{1k}+E^+_{2k}$, $E^-_{pk}+E^+_{1k}$ and
$E^-_{pk}+E^+_{2k}$ for a system with following set of parameters:
$J=0.01U$, $\gamma=0.06U$, $\Omega=0.01 U$, $\mu=0.2U$ and
$\zeta=0.4$. In this case, the system is in a superfluid with a BEC
at the momentum $[1.34 a^{-1},1.34 a^{-1}]$
where $a$ is the lattice constant. Fig.~\ref{fig5}
(a) shows $2E^{+}_{2k}$, which ranges from $0$ at BEC wavevector
(the gapless point) to $0.4U$.  All the other plots in
Fig.~\ref{fig5} show similar features except the fact that these
excitations are gapped. In Fig.~\ref{fig5} (b), $E^+_{1k}+E^+_{2k}$
ranges from $0.1 U$ to $0.42U$. $2E^+_{1k}$ ranges from $0.21 U$ to $0.51U$
as 
shown in Fig.~\ref{fig5} (c). The excitations involving band
transitions, $E^-_{pk}+E^+_{2k}$ ranges from $0.86 U$ to $1.06U$
[Fig.~\ref{fig5} (d)],  and $E^-_{pk}+E^+_{1k}$ [Fig.~\ref{fig5}
(e)]  ranges from $0.9 U$ to $1.13U$.

The optical modulation response for the above system is shown in
Fig~\ref{fig7}. Fig.~\ref{fig7}(a) shows the response at
$\omega=0.02U$, where a single contour of excitations corresponding
to exciting two phase modes, $\omega=2E^{+}_{2k}$ is seen around the
BEC wavevector $[k_0,k_0]$. Fig.~\ref{fig7}(b) shows the response at
$\omega=0.15U$, where two contours of excitations are seen, the
outer one corresponding to exciting two phase modes,
$\omega=2E^{+}_{2k}$, and the inner one corresponding to exciting a
phase and an amplitude mode $\omega=E^{+}_{2k}+E^{+}_{1k}$.
Fig.~\ref{fig7}(c) shows the response at $\omega=0.25U$, where three
contours of excitations are seen, the outermost one corresponding to
exciting two phase modes, $\omega=2E^{+}_{2k}$, the middle one
corresponding to exciting a phase and an amplitude mode
$\omega=E^{+}_{2k}+E^{+}_{1k}$, and the innermost contour
corresponding to exciting two amplitude modes, $\omega=2E^{+}_{1k}$.
As the frequency is increased to $\omega=0.42U$, the two phase modes
disappear and two contours corresponding to
$\omega=E^{+}_{2k}+E^{+}_{1k}$ and $\omega=2E^{+}_{2k}$ remain, as
seen in Fig. ~\ref{fig7}(d). As the frequency is further increased
to $\omega=0.5U$, only the contour corresponding to
$\omega=2E^{+}_{1k}$ appears in Fig.~\ref{fig7}(e).
As modulation frequency is increased further, the system shows no
response, till the frequency reaches $\omega=0.86U$. Beyond this
point, a contour of excitations corresponding to
$\omega=E^{+}_{2k}+E^{-}_{pk}$ appears around the BEC wavevector, as
seen in Fig.~\ref{fig7}(f) for $\omega=0.87U$. Beyond $\omega=0.9U$,
a second contour, $\omega=E^{+}_{1k}+E^{-}_{pk}$ appears  in
Fig.~\ref{fig7}(g). Finally  Fig.~\ref{fig7}(h) shows the response
at $\omega=1.1U$, where a single contour corresponding to
$\omega=E^{+}_{1k}+E^{-}_{pk}$ is present. The distinct pattern of
the spectrum is better visualized in Fig.~\ref{fig9} (d)-(f) where
the lattice modulation response is plotted as a function of
frequency along different radial cuts in the Brillouin zone making
angles $\theta=\pi/2$ , $\theta=\pi/4$ and $\theta=\pi/6$ with the
$k_x$ axis respectively. For the $\theta=\pi/2$ and $\theta=\pi/6$
cut, which does not pass through the condensate wavevector, all the
spectra are gapped, while the $\theta=\pi/4$ cut, which passes
through the condensate location shows the gapless Goldstone mode.
This clear qualitative distinction along various cuts can lead to a
precise location of the condensate wavevector. In addition
information about spectral weights can also be obtained from the
lattice modulation signal.




\section {Discussion}
\label{conclusion}

In this work, we have studied a system of two species of interacting
bosons in an optical lattice to modulation of the lattice potential.
The spin states of the bosons are coupled through a spin-orbit
coupling, which is implemented either by Raman dressing or by
time-dependent magnetic field gradients. This system shows a
superfluid-Mott insulator quantum phase transition as a function of
increasing interaction strength. In addition, as a function of the
relative strength of the spin-orbit coupling to the hopping
amplitude, the system in the superfluid phase shows a transition
from an ordinary superfluid phase with a BEC at the zone center to a
twisted superfluid phase with a BEC at a finite momentum. We have
provided a technique for differentiating between the different
phases of such a system based on the response of such a system on to
a modulating optical lattice.

In addition to finding the location of the precursor peaks in the MI
phase and condensate position in the SF phase which can also be
obtained by other standard experimental techniques such as
time-of-flight measurements. However, lattice modulation
spectroscopy provides several additional information. First, one can
use this technique to map out the single-particle excitation
spectrum of the bosons. In the Mott phase, it provides the effective
mass of the dispersion around the minimum gap. In the superfluid
phase, this technique not only shows the presence of gapless
excitations, a quantitative estimate of the speed of sound $v_s$
(the slope of the gapless mode) and the mass of the Higgs (gapped)
mode can be extracted from the experimental data. Further, since the
matrix elements of the lattice modulation operator can be explicitly
calculated within our technique, the spectral weight of the
different modes can also be extracted from the modulation
spectroscopy response. In this context, we would like to note that
the expression of the response in terms of the spectral function is
more general than the particular approximation used to calculate it
in this paper. For example, in general the Higgs mode will develop a
width due to decay to two phase modes. This can also be computed
from the modulation spectroscopy response. Finally, we would like to
point out that there is a technical advantage of our calculation
over its counterpart for Bragg spectroscopy. This advantage stems
from the fact that the optical modulation response, being a zero
momentum transfer process, does not receive large contribution from
the vertex correction terms. Thus, even if the poles in the single
particle Greens function are broadened due to self energy
corrections, the optical modulation response function will pick out
the frequency convolution of the one particle spectral function at
the same momentum. Thus this specific spectroscopic method provides
direct access to single particle excitations in the system.

The experimental verification of our theory involves use of standard
spectroscopy experiment techniques\cite{rf_spec,Bragg_spec}. The
specific experiment that we propose involves modulating $J$ and
$\gamma$ by a laser creating an additional optical lattice with
modulation frequency. After this modulation, one turns off the trap
and the lattice and measures the position distribution of the
outgoing bosons as done in any standard time-of-flight measurement.
The position distribution of these bosons under standard
experimental conditions \cite{blochreview} reproduces their momentum
distribution $n_{\rm mod}({\bf k},\omega)$ inside the trap (in the
presence of the modulating lattice). We suggest a comparison of
$n_{\rm mod}({\bf k},\omega)$ to the momentum distribution $n({\bf
k})$ of the bosons without the modulating lattice potential to
obtain $\delta n({\bf k},\omega) = n_{\rm mod}({\bf
k},\omega)-n({\bf k})$. We expect $\delta n({\bf k},\omega)$ to
provide the necessary information about the boson spectral function
and carry the signature of the Mott and the superfluid states as
described in Secs.\ \ref{optmod_mott} and \ref{optmod_sf}. We note
that realization of these experiments requires that the thermal
smearing of the contours would be small enough to distinguish
between the different phases. In the Mott phases this requires $k_B
T \ll 0.2 U$, where $k_B$ is the Boltzman constant and $k_B T^{\ast}
=0.2 U$ is the melting temperature of the MI phase \cite{gerbier}.
This is readily achieved in standard experiments where $U \sim 2-5\,
{\rm KHz}\, \sim 200-500\, {\rm nK}$. In the SF phase, the precise
nature of the low-energy contours for the Goldstone modes may be
difficult to discern since it would require a small $T$. However,
the amplitude modes which occurs at finite energy scale would still
be observable in a straightforward manner within current
experimental resolution. An estimate of the maximal allowed thermal
smearing for the SF phase comes from the criteria $T \ll T_c$ where
$T_c$ is the critical temperature of the SF phase which can be
estimated to be $k_B T_c \simeq z_0 {\rm Max}[J, \gamma]\simeq 0.1 U
\simeq 20-50\,{\rm nK}$. We note that $T \sim 1\, {\rm nK}$ have
been achieved in standard ultracold atom experiments \cite{gerbier}.

In conclusion, we have shown, via explicit computation of $\delta
n({\bf k},\omega)$ in the strong coupling regime, that the response
of a spin-orbit coupled Bose system to a modulated optical lattice
can differentiate between (a) Mott and superfluid phases and (b)
systems with finite momentum BEC and zero momentum BEC. The momentum
resolved nature of the optical modulation spectroscopy, which
provides information about the one particle spectral function of the
system, resolves the superfluid phase from the insulator phase by
presence/absence of gapless Goldstone modes in the two phases.
Further, the pattern of excitation contours appearing in the
Brillouin zone as the frequency is tuned is distinct in the two
phases and further helps to distinguish the phases. The momentum
resolution can also resolve states with condensates at finite
momentum from states with condensate at the zone center by looking
at the location of  the low energy excitations, which are always
centered around the momentum where the BEC forms. We have suggested
concrete experiments which can verify our theory.


\begin{thebibliography}{99}
\bibitem{ftnote2} We neglect the fact that the Raman dressed ``spin states'' are
slightly different from the original spin states.
\bibitem{ftnote1} In reality, when the deep lattice assumption is
relaxed, the co-efficients $\delta\gamma/\gamma$ and $\delta V/V$
would be slightly different. In principle one can also modulate only
the spin-orbit term by modulating the Raman lasers. These cases will
change the matrix elements in our calculations a little bit, but the
main conclusions remain unaffected.
\end{thebibliography}
\end{document}